\documentclass[11pt,a4paper]{article}
\pdfoutput=1
\usepackage{a4wide}
\usepackage{amsfonts}
\usepackage{amssymb}
\usepackage{amsmath}
\usepackage{graphicx}
\usepackage{booktabs}
\usepackage{array}
\usepackage{color}
\usepackage{ulem}
\usepackage[small,bf]{caption}
\usepackage{cite}
\usepackage[usenames,dvipsnames]{xcolor}
\usepackage{subfigure}
\usepackage{verbatim}

\def\1{\mathbf{1}}
\def\3{\mathbf{3}}
\def\2{\mathbf{2}}

\allowdisplaybreaks[1]

\numberwithin{equation}{section}

\newcounter{mysubequation}[equation]

\definecolor{pink}{rgb}{1.,.2,.8}

\begin{document}

\begin{titlepage}

\vspace*{-15mm}
\begin{flushright}
SISSA 33/2015 FISI\\
IPMU15-0118\\
TTP15-024
\end{flushright}
\vspace*{0.7cm}

\begin{center}
{ \bf\LARGE Leptogenesis in an SU(5)$\, \times \,$A$_{\mathbf{5}}$\\[1mm]
Golden Ratio Flavour Model: Addendum}
\\[8mm]
Julia Gehrlein$^{\, a,}$ \footnote{E-mail: 
\texttt{julia.gehrlein@student.kit.edu}},
S. T.\ Petcov$^{\, b,c,}$ \footnote{Also at:
 Institute of Nuclear Research and Nuclear Energy,
  Bulgarian Academy of Sciences, 1784 Sofia, Bulgaria.},
Martin Spinrath$^{\, a,}$ \footnote{E-mail: \texttt{martin.spinrath@kit.edu}},
Xinyi Zhang$^{\, d,}$ \footnote{E-mail: \texttt{xinyizhang18@gmail.com}}
\\[1mm]
\end{center}
\vspace*{0.50cm}
\centerline{$^{a}$ \it Institut f\"ur Theoretische Teilchenphysik, Karlsruhe Institute of Technology,}
\centerline{\it Engesserstra\ss{}e 7, D-76131 Karlsruhe, Germany}
\vspace*{0.2cm}
\centerline{$^{b}$ \it SISSA/INFN, Via Bonomea 265, I-34136 Trieste, Italy }
\vspace*{0.2cm}
\centerline{$^{c}$ \it Kavli IPMU (WPI), 
University of Tokyo, Tokyo, Japan}
\vspace*{0.2cm}
\centerline{$^{d}$ \it School of Physics and State Key Laboratory of Nuclear Physics and
Technology,}
\centerline{\it Peking University, 100871 Beijing, China}
\vspace*{1.20cm}

\begin{abstract}
\noindent
 We derive and discuss the solution of the Boltzmann equations for
leptogenesis in a phenomenologically viable 
SU(5)$\, \times \,$A$_{5}$ golden ratio flavour model 
proposed in \cite{Gehrlein:2014wda,Gehrlein:2015dxa}. 
The model employs, in particular, the seesaw mechanism of 
neutrino mass generation.
We find that the results on the baryon asymmetry of the Universe,
obtained earlier in \cite{Gehrlein:2015dxa} using approximate 
analytic expressions for the relevant CP violating 
asymmetry and efficiency factors, are correct, as was expected,
up to 20-30\%. The phenomenological predictions for 
the low energy neutrino observables, derived using values of the 
parameters of the model for which we reproduce the observed value of 
the baryon asymmetry, change little with respect to those 
presented in \cite{Gehrlein:2015dxa}. Among the many predictions 
of the model we find, for instance, that 
the neutrinoless double beta decay effective Majorana mass $m_{ee}$ 
lies between 3.3~meV and 14.3~meV.

\end{abstract}

\end{titlepage}
\setcounter{footnote}{0}

\section{Introduction}

The origin of flavour is one of the most challenging unresolved 
fundamental problems in particle physics.
The questions of why there are three generations (not more and not less), 
of the origin of the hierarchies of the fermion masses and of the very
different quark and neutrino mixing patterns are still far from 
having received a satisfactory explanation.

In recent years an approach to the problem of flavour 
based on discrete flavour symmetries became widely used especially 
in treating the flavour problem in the lepton sector, 
for a recent review see, e.g., \cite{King:2013eh}.
A large number of models employing discrete flavour symmetries 
have been proposed. However, many of these models focus only on
leptons and only reproduce the observed neutrino mixing angles 
with possibly a few additional predictions for the leptonic CP violation 
phases and/or the absolute neutrino mass scale.

Here we will focus on a particular model \cite{Gehrlein:2014wda} which
reproduces {\it all flavour information in the quark sector} and 
in addition to reproducing
the mixing angles in the neutrino sector, 
provides predictions for the absolute neutrino mass
scale and the leptonic CP violation phases. 
In \cite{Gehrlein:2015dxa} we discussed a slight modification 
of the original model, which allowed us to
accommodate successfully the generation of the baryon asymmetry 
of the Universe within the leptogenesis scenario \cite{Fukugita:1986hr}.
In that previous publication \cite{Gehrlein:2015dxa} 
we used analytic approximations to calculate
the baryon asymmetry, which can be expected to be correct only up to 20-30\%.
In the present article  we go beyond these approximations 
and calculate the baryon asymmetry by solving the relevant 
system of Boltzmann equations numerically. 
We show that using this more precise method of calculation of the 
baryon asymmetry one can still generate successfully the observed 
value of it in the model considered.  
We discuss also the impact of the new results 
on the baryon asymmetry on the predictions of the 
low energy observables of the model - on the 
 correlation between the angles $\theta_{13}$ and $\theta_{23}$,
on the values of the leptonic CP violation phases, 
on the value of the effective Majorana mass in neutrinoless double beta decay, 
$m_{ee}$, etc.

 The article  is organised as follows.
In Section 2 we review the model constructed in \cite{Gehrlein:2014wda} 
and its modification proposed in  \cite{Gehrlein:2015dxa}.   
In Section 3 we discuss the Boltzmann equations 
and the solutions for the baryon asymmetry we obtain.
We update the results on the neutrino masses and mixing angles 
previously obtained in  
\cite{Gehrlein:2014wda,Gehrlein:2015dxa} in Section 4. 
Section 5 contains summary and conclusions.

\section{The Leptonic Yukawa and Majorana Mass Matrices}

 In this section we briefly recapitulate the Yukawa couplings
and Majorana mass matrices in the lepton sector of the model of interest 
to fix notations. The
structure of these matrices is justified by the flavour 
symmetries of the model and is discussed
extensively in \cite{Gehrlein:2014wda, Gehrlein:2015dxa}. The interested
reader is referred to these articles for details.

The right-handed neutrino Majorana mass matrix reads
\begin{align}
 M_{\text{RR}} &= y_2^n \begin{pmatrix}
2 \sqrt{\frac{2}{3}}(v_2 + v_3) & -\sqrt{3} v_2 & -\sqrt{3} v_2 \\
-\sqrt{3}v_2 & \sqrt{6}v_3 &- \sqrt{\frac{2}{3}}(v_2 + v_3)\\
-\sqrt{3}v_2 & -\sqrt{\frac{2}{3}}(v_2 + v_3) & \sqrt{6} v_3\\
\end{pmatrix} \;
\label{eq:massmatrix_righthanded}
\end{align}
%
where $v_2$ and $v_3$ are complex (vevs) of a flavon breaking the
A$_5$ family symmetry. This matrix is of the golden ratio 
pattern type A \cite{GoldenRatioA}, i.e., it is diagonalised by
\begin{equation}
U_{\text{GR}}=
\begin{pmatrix}
	\sqrt{\frac{\phi_{g}}{\sqrt{5}}} & \sqrt{\frac{1}{\phi_{g}\sqrt{5}}} & 0\\
	-\sqrt{\frac{1}{2\phi_{g}\sqrt{5}}} & \sqrt{\frac{\phi_{g}}{2\sqrt{5}}} & \frac{1}{\sqrt{2}}\\
	\sqrt{\frac{1}{2\phi_{g}\sqrt{5}}} & -\sqrt{\frac{\phi_{g}}{2\sqrt{5}}} & \frac{1}{\sqrt{2}}
	\end{pmatrix}
	 P_{0} \;,
	 \label{eq:U_GR}
\end{equation}
%
 where  $\phi_g=\frac{1+\sqrt{5}}{2}$ is the golden ratio 
and $P_{0}$ is diagonal matrix containing the two CP violation 
Majorana phases  $\alpha_{1}$ and  $\alpha_{2}$, 
$P_{0}$ = Diag$(\text{exp}(- \tfrac{\text{i} \alpha_{1}}{2}),
\text{exp}(- \tfrac{\text{i}\alpha_{2}}{2} ),1 )$.
The phases  $\alpha_{1}/2$ and  $\alpha_{2}/2$ are related to those 
in the convention used by  the Particle Data Group \cite{Beringer:1900zz},
 $\alpha_{21}/2$ and  $\alpha_{31}/2$, as follows:
$\alpha_{21}  =\alpha_{2} - \alpha_{1}$, 
$\alpha_{31} = - \alpha_{1}$.

 The matrix of charged lepton Yukawa couplings has the form:
\begin{align}
Y_e &=  \begin{pmatrix}
0 & - 1/2 a_{21} & 0\\
6 a_{12} & 6 a_{22} & 6 a_{32}\\
0 & 0 & - 3/2 a_{33}\\
\end{pmatrix}~\; ,
\label{eq:yuk_charged}
\end{align}
%
where the $a_{ij}$ are complex parameters
which are fixed by the quark sector (since it is
a GUT model) and the charged lepton masses \cite{Gehrlein:2014wda}.
Note that we did not use here standard GUT relations but
the relations proposed in \cite{Antusch:2009gu} which are in good
agreement with the current data on fermion masses and the Higgs mass results
\cite{Antusch:2008tf, Antusch:2012fb, Antusch:2012gv}.
Since the $a_{ij}$ depend on $\tan \beta$ and to redo the fit is very time consuming
we have fixed this parameter here to 30 which is in good agreement with the
aforementioned GUT relations.

 The matrix of neutrino Yukawa couplings can be written as
\begin{equation}
Y_\nu = Y_\nu^{\text{LO}} + \delta Y_\nu \;. 
\end{equation}
%
 The matrix
\begin{equation}
 Y_\nu^{\text{LO}} = y_1^n \begin{pmatrix} 1 & 0 & 0 \\ 0 & 0 & 1 \\ 0 & 1 & 0  \end{pmatrix} \;,
 \label{eq:yukawamatrix_neutrino}
\end{equation}
%
 appeared in the original model \cite{Gehrlein:2014wda}, while
\begin{equation}
\delta Y_\nu \equiv | y_1^n | c \text{ e}^{\text{i } \gamma} \begin{pmatrix}0&1&0\\-1&0&0\\0&0&0\\ \end{pmatrix} \;,
\label{eqn:Noperator}
\end{equation}
%
was introduced in the second study \cite{Gehrlein:2015dxa}.
The phases can be chosen such that $y_1^n$ is real and then it turned
out that the baryon asymmetry is generated only by the small correction
$\delta Y_\nu$ (note that $c \ll 1$). For more details and analytical estimates
the reader is referred 
to \cite{Gehrlein:2015dxa}.

 It proves convenient for our further discussion to define 
the parameters which will play an important role in the analysis 
we will perform:
 \begin{align}
M_1&=\frac{1}{\sqrt{6}}(X+Y)=\frac{1}{\sqrt{6}}|X||1+\rho \text{ e}^{\text{i}\phi}|\text{e}^{\text{i}\phi_1}, ~\phi_1=\text{arg}(X+Y)~,\\
M_2&=\frac{1}{\sqrt{6}}(X-Y)=\frac{1}{\sqrt{6}}|X||1-\rho \text{ e}^{\text{i}\phi}|\text{e}^{\text{i}\phi_2},~ \phi_2=\text{arg}(X-Y)~,\\
M_3&=\sqrt{\frac{2}{3}} X=\sqrt{\frac{2}{3}} |X|\text{ e}^{\text{i}\phi_3},~ \phi_3=\text{arg}(X)~,
\end{align}
where 
\begin{align}
X&=(4 v_3+ v_2)y_2^n~,\\
Y&=3\sqrt{5}v_2 y_2^n~,\\
\rho&=\left|\frac{Y}{X}\right|~,\\
\phi&=\text{arg}(Y)-\text{arg}(X)~.
\end{align}
One of the Majorana phases which we choose to be $\phi_1$ can be set to zero by
applying a redefinition of the heavy Majorana fields.
The remaining two phases $\phi_2$ and $\phi_3$ can be expressed in terms of
$\rho$ and $\phi$ using the complex mass sum rule $M_1+M_2=M_3$
\begin{align}
\cos\phi_2&=\frac{|M_3|^2-|M_1|^2-|M_2|^2}{2|M_1||M_2|}=\frac{1-\rho^2}{\sqrt{1-2\rho^2\cos 2\phi+\rho^4}}~,\label{alphaphi2}\\
\cos\phi_3&=\frac{|M_1|^2-|M_2|^2+|M_3|^2}{2|M_1||M_3|}=\frac{1+\rho \cos\phi}{\sqrt{1+2\rho\cos \phi+\rho^2}}~.\label{alphaphi3}
\end{align}
%
We note that only normal ordering is viable in the model considered 
\cite{Gehrlein:2014wda, Gehrlein:2015dxa},
and the Yukawa couplings are degenerate in LO, 
so that we have $|M_3|<|M_2|<|M_1|$.
The Majorana phases $\alpha_1$, $\alpha_2$ are employed in the 
 renormalisation group evolution package
REAP \cite{REAP} we are going to use in our numerical analysis, 
and the phases $\phi_2$ and $\phi_3$ are related, 
 up to corrections of order $c^2$, 
via
\begin{align} 
\label{eq:MajoranaPhases}
 \alpha_1 = -\phi_3 \text{ and } \alpha_2 = \phi_2 - \phi_3\,. 
\end{align}

\section{Boltzmann Equations}
\label{sec:BE}

In this section we discuss the Boltzmann equations for this model.
Since we set the leptogenesis scale to be  $M_S\approx10^{13}$ GeV,
we find with $\tan \beta=30$, $10^9 (1+\tan^2\beta)$ GeV$<M_S<10^{12} (1+\tan^2\beta)$~GeV. Values of $M_S$ in this interval correspond  \cite{Pascoli:2006ci} to 
the two-flavour leptogenesis regime \cite{Nardi:2006fx,Abada:2006fw}. 
We will perform the analysis 
of the baryon asymmetry generation in this regime.

 We use the set of Boltzmann equations in supersymmetric 
leptogenesis \cite{Antusch:2006cw,Fong:2010qh,Meroni:2012ze}
which we briefly summarise below (for notational details and 
further explanations, see the original papers).

 The baryon asymmetry generated in the two-flavour 
regime in leptogenesis is determined, 
in particular, by the 
evolution of the heavy Majorana neutrino and sneutrino 
number densities (abundances), $Y_{N_i}$ and $Y_{\widetilde{N}_i}$, 
and of the lepton charge and CP violating asymmetries
in the charges $L_e + L_\mu$ and $L_\tau$,  
$\hat{Y}_{\Delta_2}\equiv\hat{Y}_{\Delta_e}+\hat{Y}_{\Delta_\mu}$
and $\hat{Y}_{\Delta_\tau}$ (where $\hat{Y}_{\Delta_{\ell}}\equiv Y_{\Delta_\ell}+ Y_{\Delta_{\tilde{\ell}}}$, 
with $\Delta_{\ell(\tilde{\ell})}\equiv B/3-L_{\ell(\tilde{\ell})}$), 
during the epoch of the evolution of the Universe 
when the abundances $Y_{N_i}$ and $Y_{\widetilde{N}_i}$ 
start to deviate from their equilibrium values 
(the out-of-equilibrium Sakharov condition \cite{Sakharov:1967dj}).
The evolution of the quantities of interest
in the epoch of interest  can be described by 
a system of coupled Boltzmann equations, 
which in the case of the MSSM and of the two-flavour regime 
we consider  read~\cite{Antusch:2006cw,Fong:2010qh,Meroni:2012ze} 

\begin{eqnarray}
	\frac{d Y_{N_i}}{dz} & = & 
- \frac{z}{s H(M_{\text{FLG}})}\,\,2\,\left(\gamma_{D}^i\,+ 
\,\gamma_{S,\,\Delta L=1}^i  \right)\,
\left(\frac{Y_{N_i}}{Y^{\rm eq}_{N_i}}\,-1\right)\,,
\label{YNNBE}\\
	\frac{d Y_{\widetilde{N}_i}}{dz} & = & 
- \frac{z}{s H(M_{\text{FLG}})}\,2\,\left(\gamma_{D}^{\tilde{i}}\,+ 
\,\gamma_{S,\,\Delta L=1}^{\tilde{i}}  \right)
\,\left(\frac{Y_{\widetilde{N}_i}}{Y^{\rm eq}_{\widetilde{N}_i}}\,-1\right)\,,\\
\frac{d\, \hat{Y}_{\Delta_2}}{dz} & = & - \frac{z}{s H(M_{\text{FLG}})}\,\sum\limits_{i=1}^3\,\left[\, \left(\epsilon_{i}^e+\epsilon_{i}^{\tilde{e}}+\epsilon_{i}^\mu+\epsilon_{i}^{\tilde{\mu}}\right)\left(\gamma_{D}^i+\gamma_{S,\,\Delta L=1}^i  \right)
\left(\frac{Y_{N_i}}{Y^{\rm eq}_{N_i}}-1\right)\, \right.\nonumber\\
&& \left.+ 
\left(\epsilon_{i}^e+\epsilon_{i}^{\tilde{e}}+\epsilon_{i}^\mu+\epsilon_{i}^{\tilde{\mu}}\right)\left(\gamma_{D}^{\tilde{i}}+\gamma_{S,\,\Delta L=1}^{\tilde{i}}  \right)
	\left(\frac{Y_{\widetilde{N}_i}}
{Y^{\rm eq}_{\widetilde{N}_i}}-1\right)\, \right.\nonumber\\
&& -\left(\,\frac{\gamma_D^{i,e}\,+\,\gamma_D^{i,\tilde{e}}+\gamma_D^{i,\mu}\,+\,\gamma_D^{i,\tilde{\mu}}}{2}
	\,+\,\gamma_{W,\,\Delta L=1}^{i,e}\,+\,\gamma_{W,\,\Delta L=1}^{i,\tilde{e}}\,+\,\gamma_{W,\,\Delta L=1}^{i,\mu}\,+\,\gamma_{W,\,\Delta L=1}^{i,\tilde{\mu}}\,\right.\nonumber\\
	&&\left.+\,\frac{\gamma_D^{\tilde{i},e}\,+\,\gamma_D^{\tilde{i},\tilde{e}}+\gamma_D^{\tilde{i},\mu}\,+\,\gamma_D^{\tilde{i},\tilde{\mu}}}{2}
	\,+\,\gamma_{W,\,\Delta L=1}^{\tilde{i},e}\,+\,\gamma_{W,\,\Delta L=1}^{\tilde{i},\tilde{e}}   \,	\,+\,\gamma_{W,\,\Delta L=1}^{\tilde{i},\mu}\,+\,\gamma_{W,\,\Delta L=1}^{\tilde{i},\tilde{\mu}}   \,\right)\, \nonumber\\
	&& \left.\,\frac{A_{22}\,\hat{Y}_{\Delta_2}+A_{2\tau}\,\hat{Y}_{\Delta_{\tau}}}{\hat{Y}_\ell^{\rm eq}}\,\right]\,,\label{YDeltaell}\\
	\frac{d\, \hat{Y}_{\Delta_\tau}}{dz} & = & - \frac{z}{s H(M_{\text{FLG}})}\,\sum\limits_{i=1}^3\,\left[\, \left(\epsilon_{i}^\tau+\epsilon_{i}^{\tilde{\tau}}\right)\left(\gamma_{D}^i+\gamma_{S,\,\Delta L=1}^i  \right)
	\left(\frac{Y_{N_i}}{Y^{\rm eq}_{N_i}}-1\right)\, \right.\nonumber\\
	&& \left.+ 
	\left(\epsilon_{\tilde{i}}^\tau+\epsilon_{\tilde{i}}^{\tilde{\tau}}\right)\left(\gamma_{D}^{\tilde{i}}+\gamma_{S,\,\Delta L=1}^{\tilde{i}}  \right)
		\left(\frac{Y_{\widetilde{N}_i}}
	{Y^{\rm eq}_{\widetilde{N}_i}}-1\right)\, \right.\nonumber\\
	&& \left.-\left(\frac{\gamma_D^{i,\tau}\,+\,\gamma_D^{i,\tilde{\tau}}}{2}
		\,+\,\gamma_{W,\,\Delta L=1}^{i,\tau}\,+\,\gamma_{W,\,\Delta L=1}^{i,\tilde{\tau}}\,+\,\frac{\gamma_D^{\tilde{i},\tau}\,+\,\gamma_D^{\tilde{i},\tilde{\tau}}}{2}
		\,+\,\gamma_{W,\,\Delta L=1}^{\tilde{i},\tau}\,+\,\gamma_{W,\,\Delta L=1}^{\tilde{i},\tilde{\tau}}   \right)\, \right.\nonumber\\
		&& \left.
		\,\frac{A_{\tau 2}\,\hat{Y}_{\Delta_2}+A_{\tau\tau}\,\hat{Y}_{\Delta_{\tau}}}{\hat{Y}_\ell^{\rm eq}}\,\right]\,,
\end{eqnarray}
%
 
In these expressions, $Y_{N_i}^{\text{(eq)}}$
is the $N_i$ (equilibrium) abundance, 
$z\equiv M_{\text{FLG}}/T$ with $M_{\text{FLG}}$ being
the mass scale of flavoured leptogenesis and $T$ being the 
temperature of the thermal bath,
$s=g_*2\pi^2T^3/45$ is the entropy density, 
$H(T) \simeq 1.66\sqrt{g_*}T^2/m_{\text{Pl}}$
is the expansion rate of the Universe, $g_*=228.75$ and the Planck mass
$m_{\text{Pl}}\simeq 1.22\times 10^{19}$~GeV. 
$\gamma_D^i$ is the thermally averaged $N_i$ decay rate,
$\gamma_{S,\Delta L=1}^i$ is the $\Delta L=1$ scattering rate 
of $N_i$ with leptons, quarks and
gauge bosons, $\gamma_D^{i,l(\tilde{l})}$ is the flavour 
dependent $N_i$ inverse decay rate,
and $\gamma_{W, \Delta L=1}^{i, l(\tilde{l})}$ is the washout 
rate of the $\Delta L=1$ scatterings.
The $A$ matrix that relates the $B/3-L_l$ asymmetry $Y_{\Delta_l}$ 
and the lepton charge asymmetry in doublets, $Y_l$, defined in $Y_l=\sum_{l^\prime} A_{ll^\prime} Y_{\Delta_{l^\prime}}$, in MSSM and the 
two-flavour regime reads \cite{Antusch:2006cw}
\begin{eqnarray}
A= \frac{1}{761} \begin{pmatrix}
-541 & 152\\
46 & -494
\end{pmatrix} \;.
\end{eqnarray}
%
 The expressions for the CP violating asymmetries generated 
in the heavy Majorana (s)neutrino decays are the same as 
those derived in \cite{Gehrlein:2015dxa}:

\begin{align}
\epsilon_1^\tau&=\frac{c \left({y_1^n}\right)^2}{8 \pi}\frac{1}{\sqrt{10}} \left( \sin\gamma \cos{\phi_2} f(\tfrac{m_1}{m_2})-\sin\gamma \frac{m_2^2}{m_2^2-m_1^2}
+\cos\gamma \sin{\phi_3} f(\tfrac{m_1}{m_3}) \right)~, \\
\epsilon_2^\tau&=\frac{c \left({y_1^n}\right)^2}{8 \pi}\frac{1}{\sqrt{10}}\left( -\sin\gamma \cos{\phi_2} f(\tfrac{m_2}{m_1})+ \sin\gamma \frac{m_1^2}{m_1^2-m_2^2}
-\cos\gamma\sin{(\phi_3-\phi_2)} f(\tfrac{m_2}{m_3}) \right)~, \\
\epsilon_3^\tau&=\frac{c \left({y_1^n}\right)^2}{8 \pi}\frac{1}{\sqrt{10}}\cos\gamma \left( -\sin{\phi_3} f(\tfrac{m_3}{m_1})+ \sin{(\phi_3-\phi_2)} f(\tfrac{m_3}{m_2}) \right)~.
\end{align}
%
Due to the fact that leading order neutrino Yukawa coupling is unitary (except for an overall factor
$\left({y_1^n}\right)^2$), we have $\epsilon_i^2\equiv\epsilon_i^e+\epsilon_i^\mu=-\epsilon_i^\tau$
to leading order.

We do not include here the thermal corrections to the CP asymmetries since they give only
negligible contributions within the  low temperature regime~\cite{Giudice:2003jh,Davidson:2008bu}.
 We neglect also the $\Delta L=2$ process for the same reason as stated 
in \cite{Gehrlein:2015dxa}, i.e., 
 given the  value of $y_1^n$,
the $\Delta L=2$ processes do not have a significant impact at 
the leptogenesis scale of interest $M_S\cong 10^{13}$ GeV.

The final baryon asymmetry is
\begin{eqnarray}
Y_B=\frac{10}{31}\left(\hat{Y}_{\Delta_2}+\hat{Y}_{\Delta_\tau}\right).
\end{eqnarray}

\begin{figure}
\centering
\includegraphics[scale = 0.7]{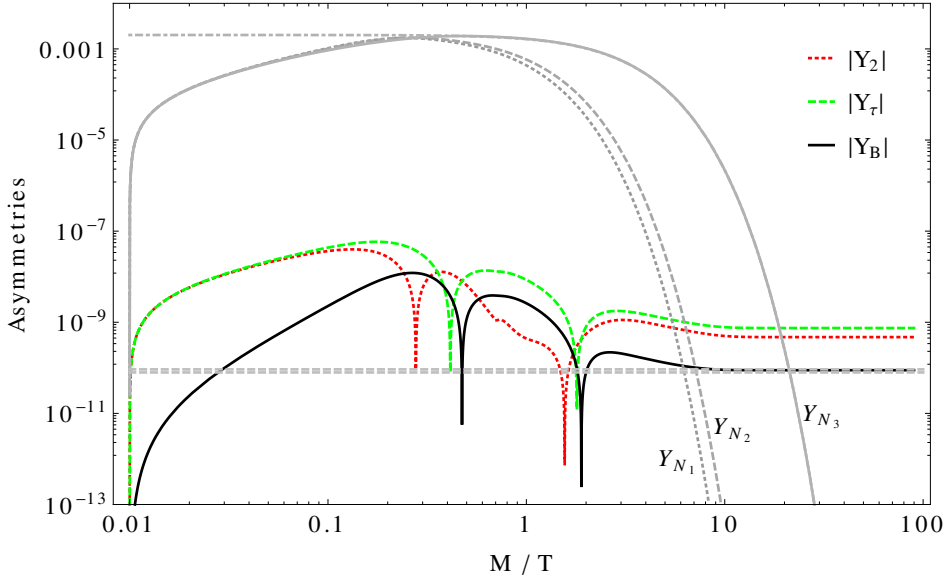}
\caption{
 Solutions for the Boltzmann equations using the example input set. The horizontal grey band represents the 3$\sigma$ region for the observed baryon asymmetry $Y_B=(8.58\pm0.22)\times 10^{-11}$, 
where, for simplicity,
we obtained the 3$\sigma$ region by multiplying the 1$\sigma$ error 
by a factor of three. 
The dot-dashed line is $Y_{N_3}^{\text{eq}}$. See text for further details.
}
\label{fig:BS5_WS_1}
\end{figure}

As an example we show the result of the Boltzmann equations in Fig.~\ref{fig:BS5_WS_1}
for a single set of parameters chosen from the numerical scan in the next section. The parameters
for this plot are
\begin{eqnarray}
\tan\beta=30,~M_3=8.51\times 10^{12} \text{ GeV},~
c=0.053,~\rho=6.29,~\gamma=3.33 \;.
\end{eqnarray}
And we have
\begin{align}
Y_B &= 8.86\times 10^{-11}~,\\
Y_{\Delta_2} &= -4.72 \times 10^{-10}~,\\
Y_{\Delta_\tau} &= 7.47 \times 10^{-10}~.
\end{align}

Note that at this point we had given two different example points in our
previous publication \cite{Gehrlein:2015dxa}. But with the improved calculations
none of them is in good agreement with the experimental result on $Y_B$ anymore
and hence we have chosen here a different example point. This shows furthermore
the importance of the current study.

\section{Results}

In this section we show the results for the masses and 
the mixing angles as well as the results
for $Y_B$ in dependence on the parameters in our model. 
To obtain realistic values for $Y_B$ coming from the 
numerical solution of the Boltzmann equations
given in Sec.~\ref{sec:BE} we had to slightly increase 
the value of the neutrino Yukawa coupling from
$y_1=0.1$ to $y_1=0.12$ which is a $20\%$ change as anticipated.

\subsection{Masses and mixing angles}
For our numerical scan we use the method as described 
in \cite{Gehrlein:2014wda}. For the
parameters which characterise the charged lepton and 
quark sector we use the fit results given
there with $\tan \beta=30$ and $M_{\text{SUSY}}=1$ TeV. 
The renormalisation group evolution
is done using the REAP package \cite{REAP}.
\begin{table}
\centering
\begin{tabular}{lcc} 
\toprule
Parameter & Best-fit ($\pm 1\sigma$) & $ 3\sigma$ range\\ 
\midrule 
$\theta_{12}$ in $^{\circ}$ & $ 33.48^{+0.78}_{-0.75}$& $31.29\rightarrow 35.91$\\[0.5 pc]
$\theta_{13}$ in $^{\circ}$ & $ 8.50^{+0.20}_{-0.21}\oplus 8.51^{+0.20}_{-0.21} $& $7.85\rightarrow 9.10 \oplus 7.87\rightarrow 9.11$\\[0.5 pc]
$\theta_{23}$ in $^{\circ}$ & $ 42.3^{+3.0}_{-1.6}\oplus 49.5^{+1.5}_{-2.2}$ & $38.2\rightarrow 53.3 \oplus 38.6\rightarrow 53.3$\\[0.5 pc]
$\delta$  in $^{\circ}$&$251^{+67}_{-59}$&$0\rightarrow 360$\\
\midrule
$\Delta m_{21}^{2}$ in $10^{-5}$~eV$^2$ & $7.50^{+0.19}_{-0.17}$ & $7.02\rightarrow 8.09$\\[0,5 pc]
$\Delta m_{31}^{2}$ in $10^{-3}$~eV$^2$~(NH) &$2.457^{+0.047}_{-0.047}$&$2.317\rightarrow 2.607$\\[0,5 pc]
$\Delta m_{32}^{2}$ in $10^{-3}$~eV$^2$~(IH) &$-2.449^{+0.048}_{-0.047}$&$-2.590\rightarrow -2.307$\\
\bottomrule
\end{tabular}
\caption{The best-fit values and the 3$\sigma$ ranges for the parameters taken from~\cite{Gonzalez-Garcia:2014bfa}. The two minima for both $\theta_{13}$ and 
$\theta_{23}$ correspond to normal and inverted mass ordering, respectively.}
\label{tab:exp_parameters}
\end{table}

There are six free parameters left in our model, the moduli $|X|$ and $|Y|$, the
phases $\phi$ and $\delta_{12}^e$ in the leading order matrices and the modulus
$|c|$ and the phase $\gamma$, which come from the correction 
to the neutrino Yukawa matrix.
We performed a random scan over these parameters and impose 
the experimental ranges for the mixing angles, mass squared differences 
(cf.~Tab.~\ref{tab:exp_parameters}) and $Y_B$ as constraints.
For $Y_B$ we used \cite{Ade:2013zuv, Bennett:2012zja}
\begin{equation}
Y_B = (8.58 \pm 0.22) \times 10^{-11} \;,
\end{equation}
%
where the 3$\sigma$ uncertainty is obtained, for simplicity, 
by multiplying the 1$\sigma$ error by a factor of three.
In order to calculate $Y_B$ we solved the Boltzmann 
equations given in Sec.~\ref{sec:BE} numerically.

Before we present the results for the normal ordering of the neutrino masses
 we comment briefly on the case of inverted ordering.
In the original model \cite{Gehrlein:2014wda} the inverted ordering was not
viable due to incompatible constraints for $\theta_{12}$ coming from the
mass sum rule on one hand and from the angle sum rule in our model on the
other hand. Due to the correction for the neutrino Yukawa matrix the mixing
angles are modified with corrections of order $c$ which nevertheless have to
be of the order of $c\approx 0.4$ to save the inverted ordering as we have
shown in an estimate given in \cite{Gehrlein:2015dxa}.
Since $c$ is associated with a higher dimensional operator such high
values are not plausible.

\begin{figure}
\centering
\includegraphics[scale=0.49]{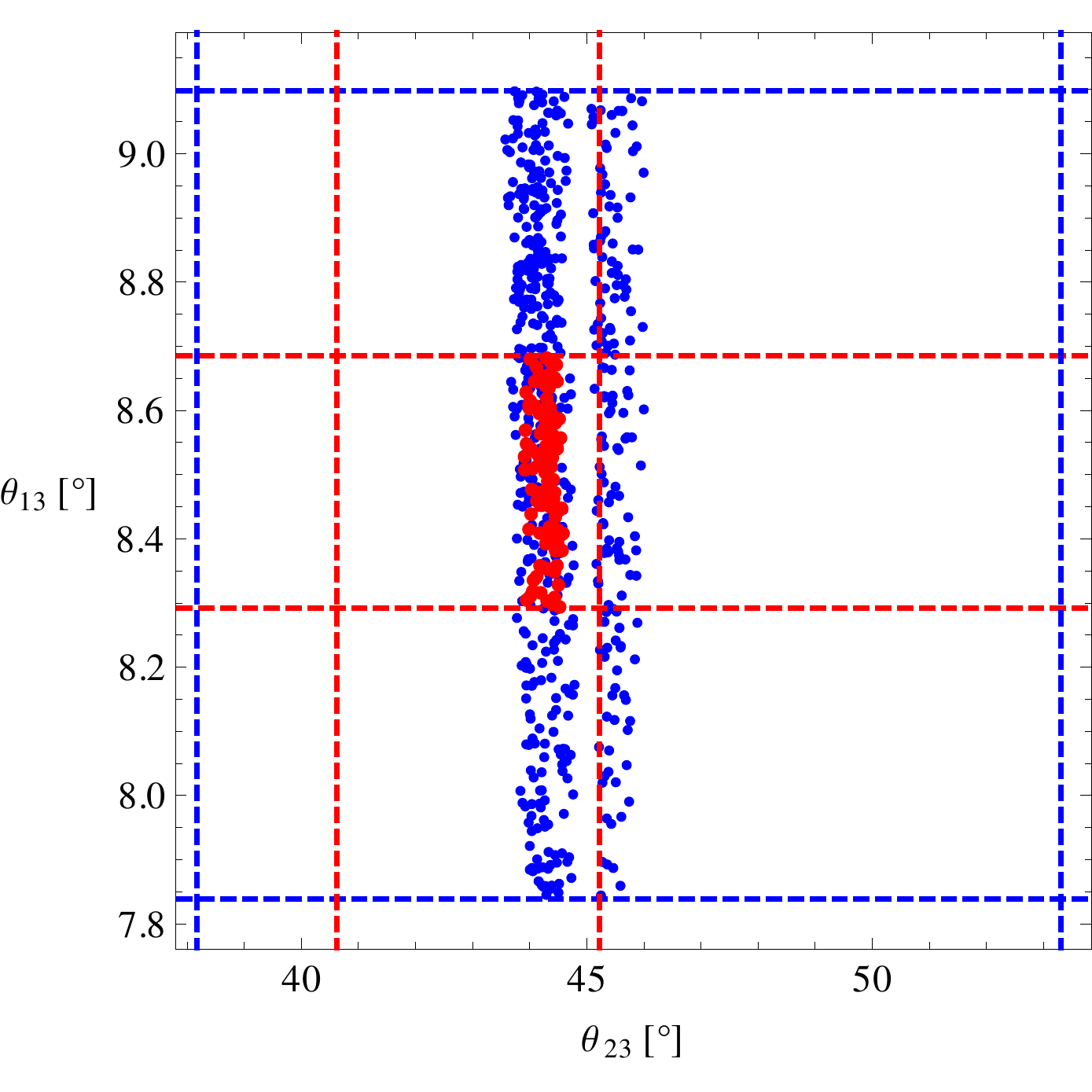} \hspace{0.7cm}
\includegraphics[scale=0.49]{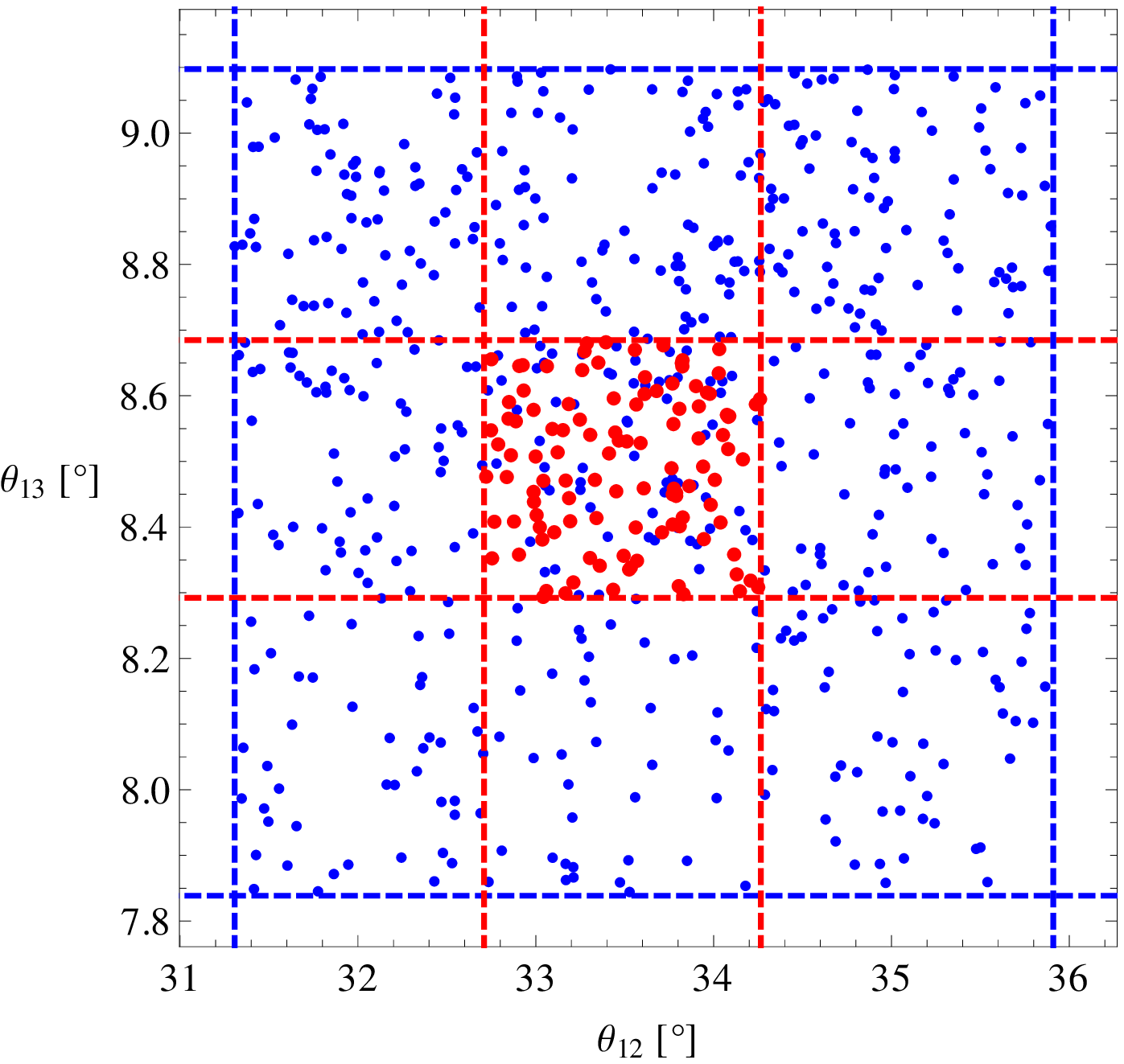}
\includegraphics[scale=0.49]{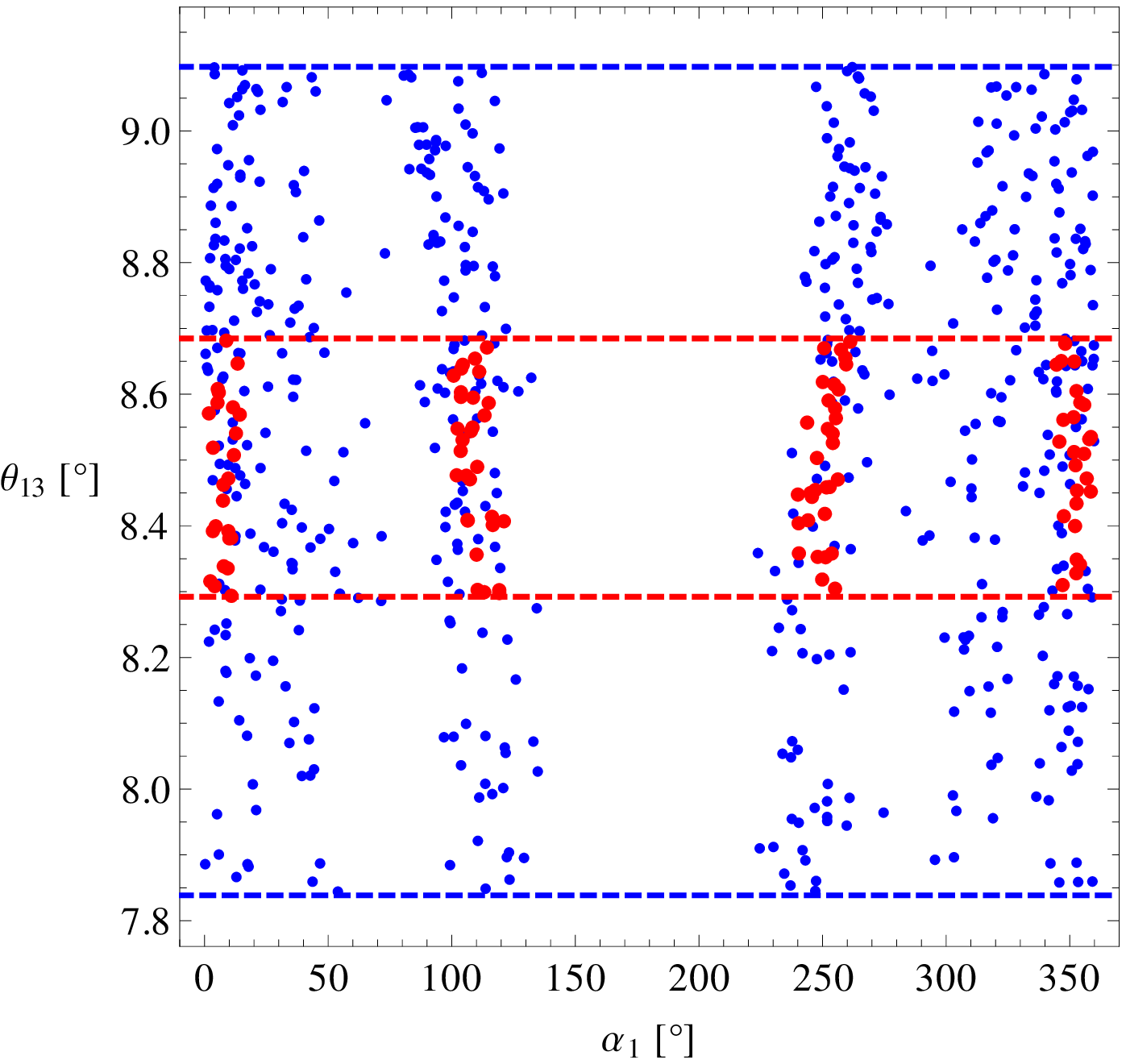} \hspace{0.7cm}
\includegraphics[scale=0.49]{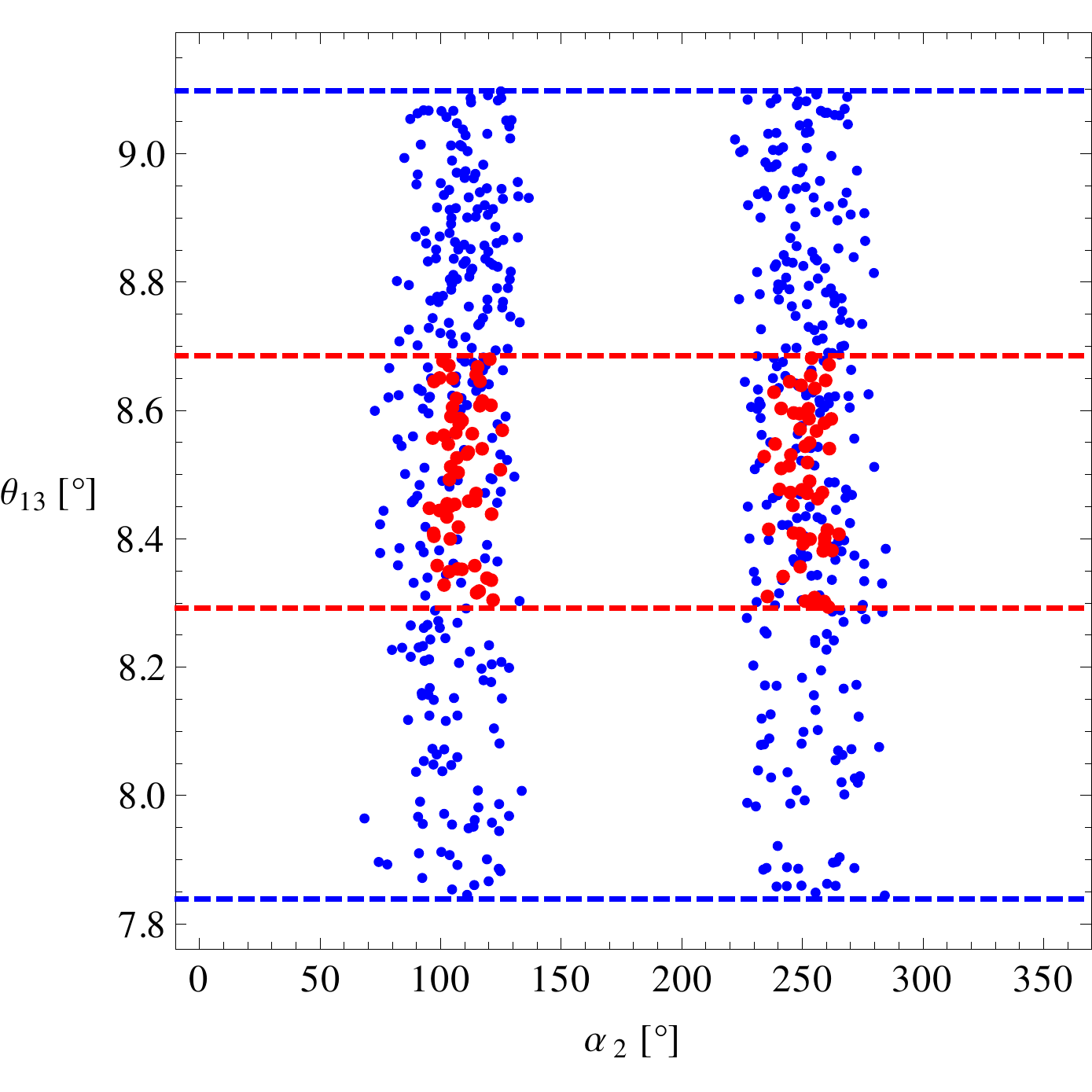}
\includegraphics[scale=0.49]{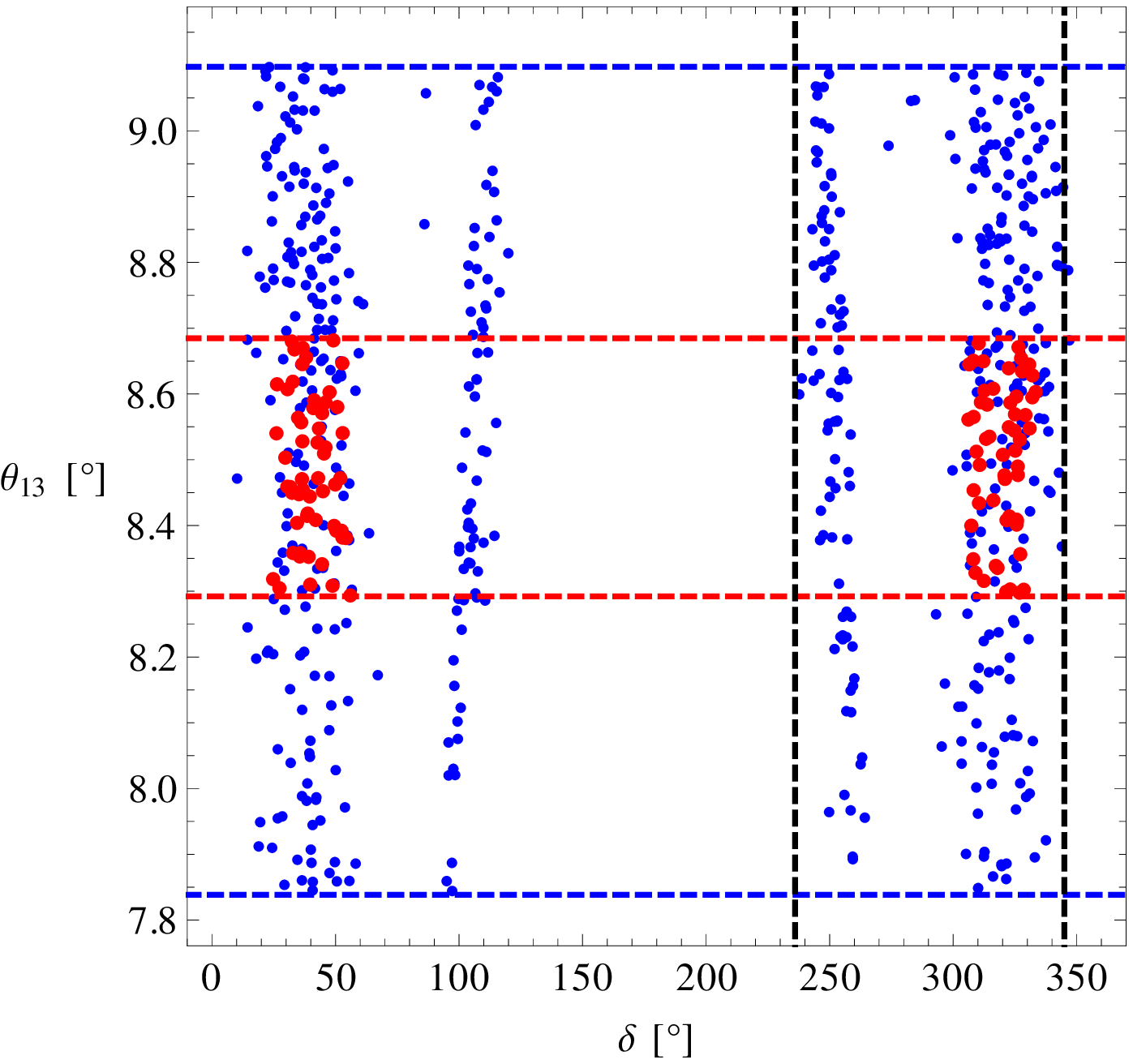} \hspace{0.7cm}
\includegraphics[scale=0.49]{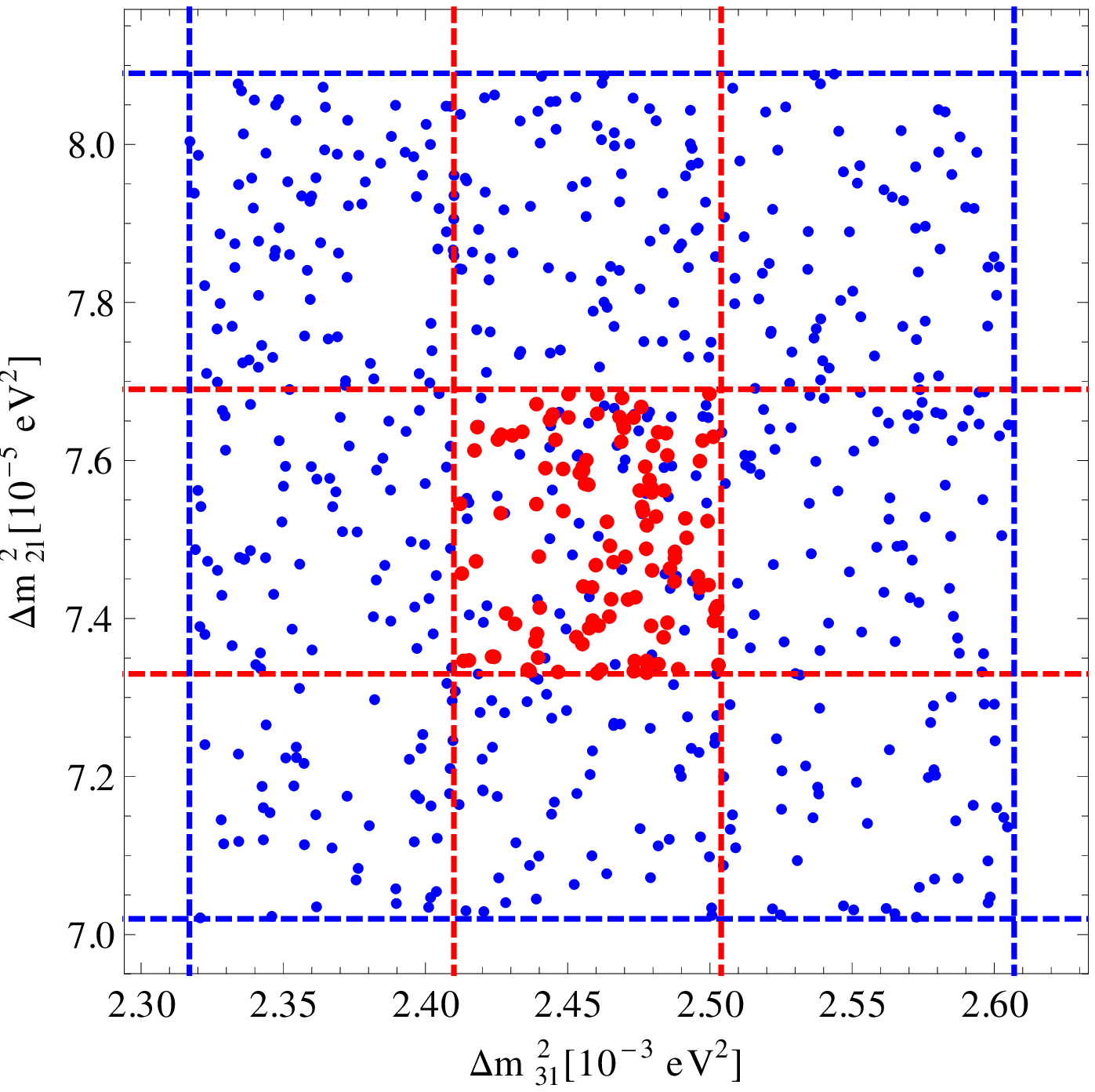}
\caption{
Results of our numerical parameter scan.
Blue (red) points are in agreement within 3$\sigma$ (1$\sigma$) of the low energy
neutrino masses and mixings and $Y_B$ in our model.
The allowed experimental 3$\sigma$ (1$\sigma$) regions are
limited by blue (red) dashed lines. The black dashed lines represent the 1$\sigma$
range for the not directly measured CP phase $\delta$ from the global
fit \cite{Gonzalez-Garcia:2014bfa}. 
}
\label{fig:mixing_parameters}
\end{figure}

Turning now to the normal ordering, we show the results for the masses and
mixing angles in Fig.~\ref{fig:mixing_parameters}. We find all parameters in
agreement with the experimental 1$\sigma$ (3$\sigma$) ranges for the masses, mixing angles and $Y_B$.

The correlations between $\theta_{13}$ and the Majorana phases are weaker than in \cite{Gehrlein:2015dxa} which is due to the constraints on the phase $\phi$ and the value of $|Y|$ coming from the numerical solution of $Y_B$. Furthermore we need here a larger value of $c$ which washes the correlations out and enlarges the ranges for the phases.
Nevertheless we find the phases to be in similar ranges as in \cite{Gehrlein:2015dxa}, namely we obtain
\begin{align}
\delta &\in [9 ^{\circ}, 119^{\circ}] ~\text{or}~ [239 ^{\circ}, 344^{\circ}] \;, \\
\alpha_1 &\in [0^{\circ}, 134^{\circ}] ~\text{or}~ [220^{\circ}, 360^{\circ}] \;, \\
\alpha_2 &\in [66 ^{\circ}, 134^{\circ}] ~\text{or}~ [222^{\circ}, 282^{\circ}] \;. 
\end{align}

\begin{figure}
\centering
\includegraphics[scale=0.55]{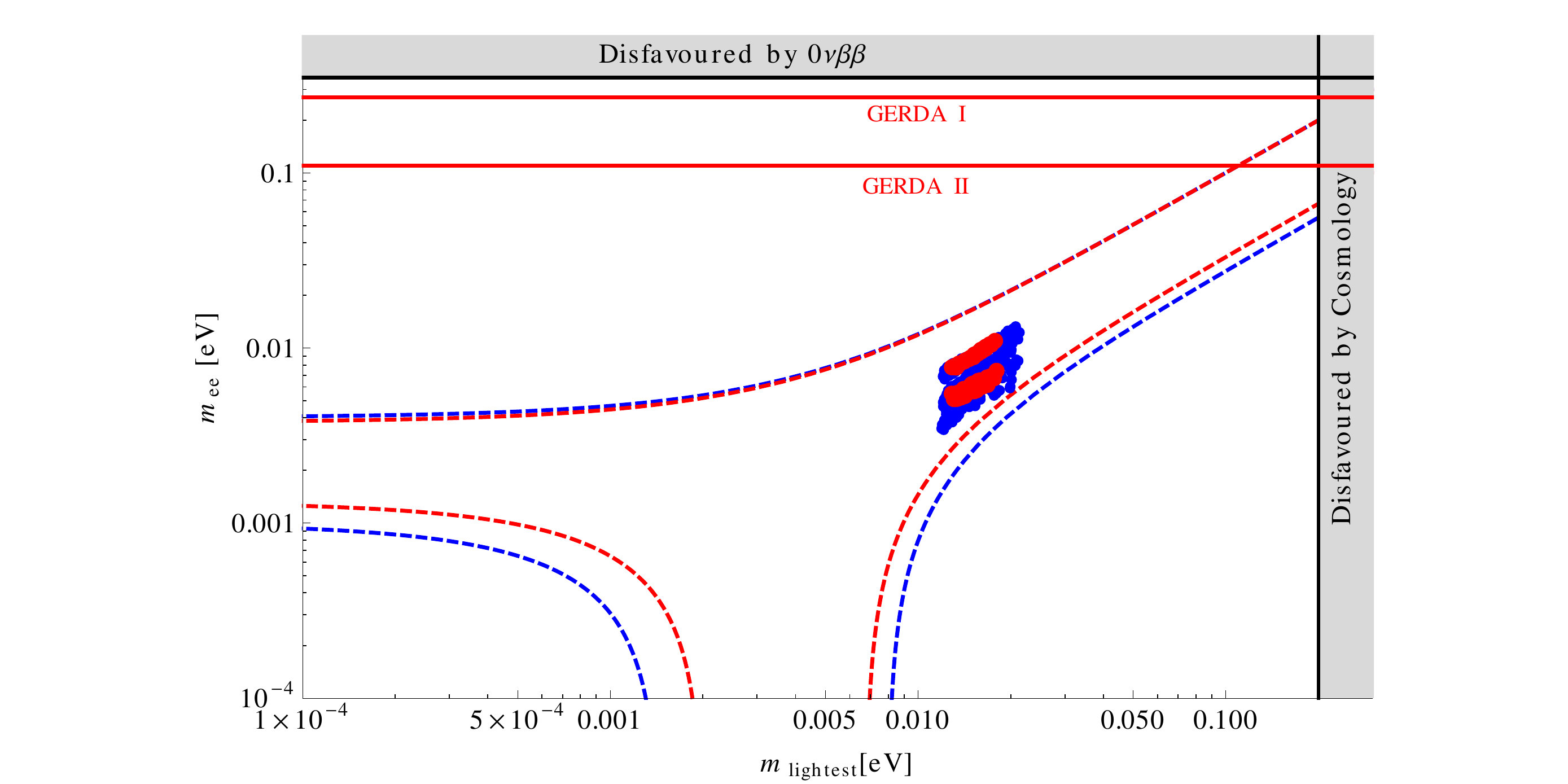}
\caption{
Prediction for the effective neutrino mass $m_{ee}$ accessible in neutrinoless double beta decay
experiments as a function of the lightest neutrino mass $m_{1}$.
The allowed experimental 3$\sigma$ (1$\sigma$) regions for the masses and mixing angles
in the case of normal ordering are limited by blue (red) dashed lines.
Blue (red) points are in agreement within 3$\sigma$ (1$\sigma$) of the low energy
neutrino masses and mixings and $Y_B$ in our model.
The grey region on the right side shows the bounds on the
lightest mass from cosmology \cite{Ade:2013zuv} and the grey region in the upper part displays the upper
bound on the effective mass from the EXO experiment \cite{Albert:2014awa}.
The red, straight lines represent the sensitivity of GERDA phase I respectively GERDA phase II \cite{Smolnikov:2008fu}.
}
\label{fig:mee_plot}
\end{figure}

 For the rephasing invariant $J_{\text{CP}}$ which determines the magnitude of 
CP violation effects in neutrino oscillations \cite{PKSP3nu88} we find
values in the ranges $ J_{\text{CP}} = \pm(0.006, 0.036)$. Our predictions 
for neutrinoless double beta decay are shown in Fig.~\ref{fig:mee_plot}.
For the lightest neutrino mass $m_1$ which is mostly determined by the
mass sum rule we obtain values between 12~meV and 22~meV. For the
observable in neutrinoless double beta decay $m_{ee}$ we obtain  values
between 3.3 meV and 14.3 meV.
For the sum of the neutrino masses we predict 
\begin{equation}
\sum m_{\nu} \in {0.077- 0.099} ~\text{eV} \;,
\end{equation}
which might be determined, e.g.,  from cosmology. So far
there is only an upper bound  \cite{Ade:2013zuv}
\begin{equation}
\sum m_{\nu} < 0.23 \text{ eV,}
\end{equation} 
which is well in agreement with our prediction.
The second observable is the kinematic mass $m_{\beta}$  as measured
in the KATRIN experiment \cite{Angrik:2005ep}
which is given as
\begin{equation}
m_{\beta}^{2}=m_{1}^{2}c_{12}^{2}c_{13}^{2}+m_{2}^{2}s_{12}^{2}c_{13}^{2}+m_{3}^{2}s_{13}^{2} \;.
\label{eq:katrin}
\end{equation}
Here we predict $m_{\beta} \approx {0.015-0.023}$~eV which is  below the
projected reach of $m_{\beta} > 0.2$~eV of KATRIN.

\subsection{Leptogenesis}
\begin{figure}
\centering
\includegraphics[scale=0.49]{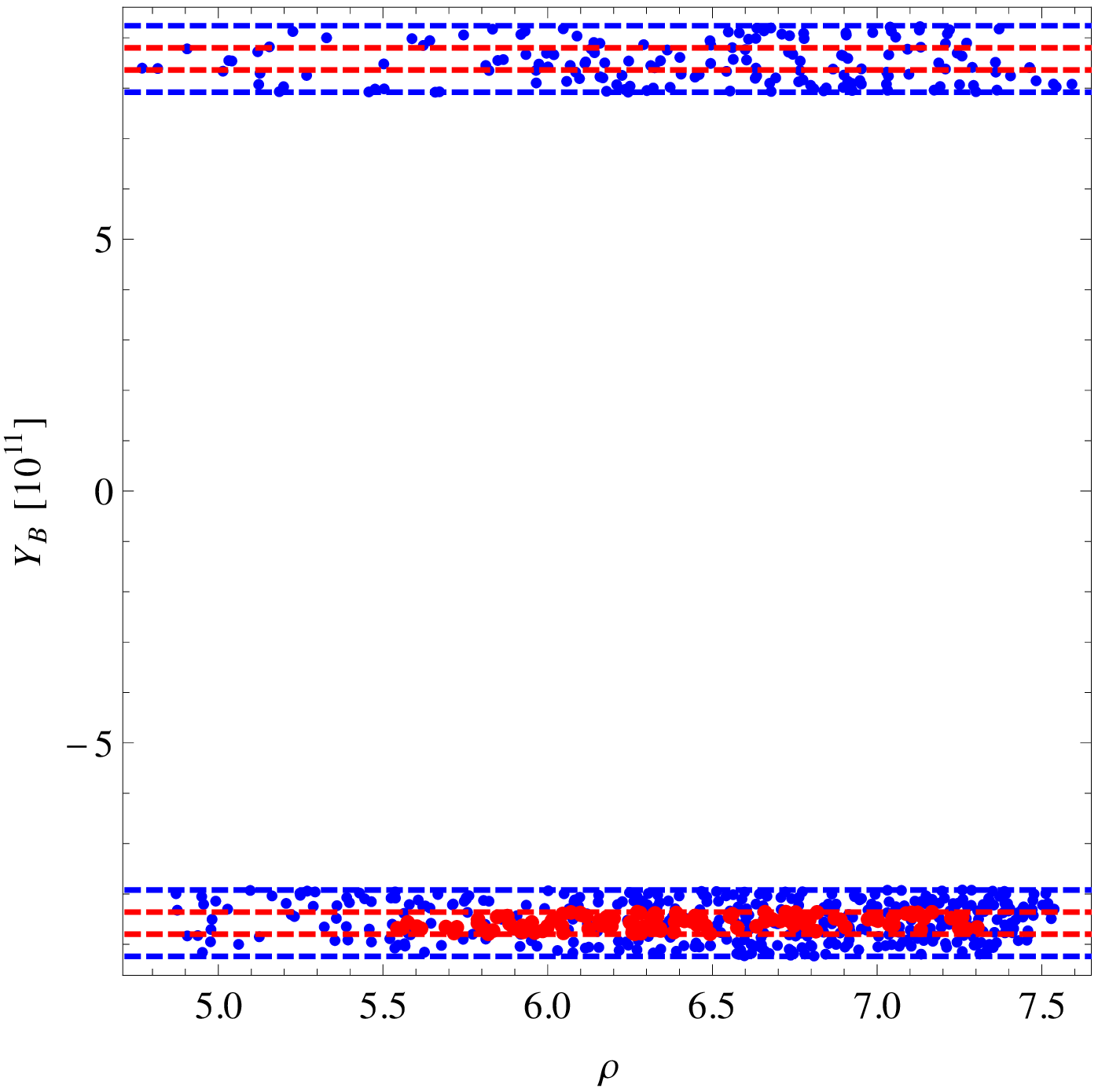} \hspace{0.7cm}
\includegraphics[scale=0.49]{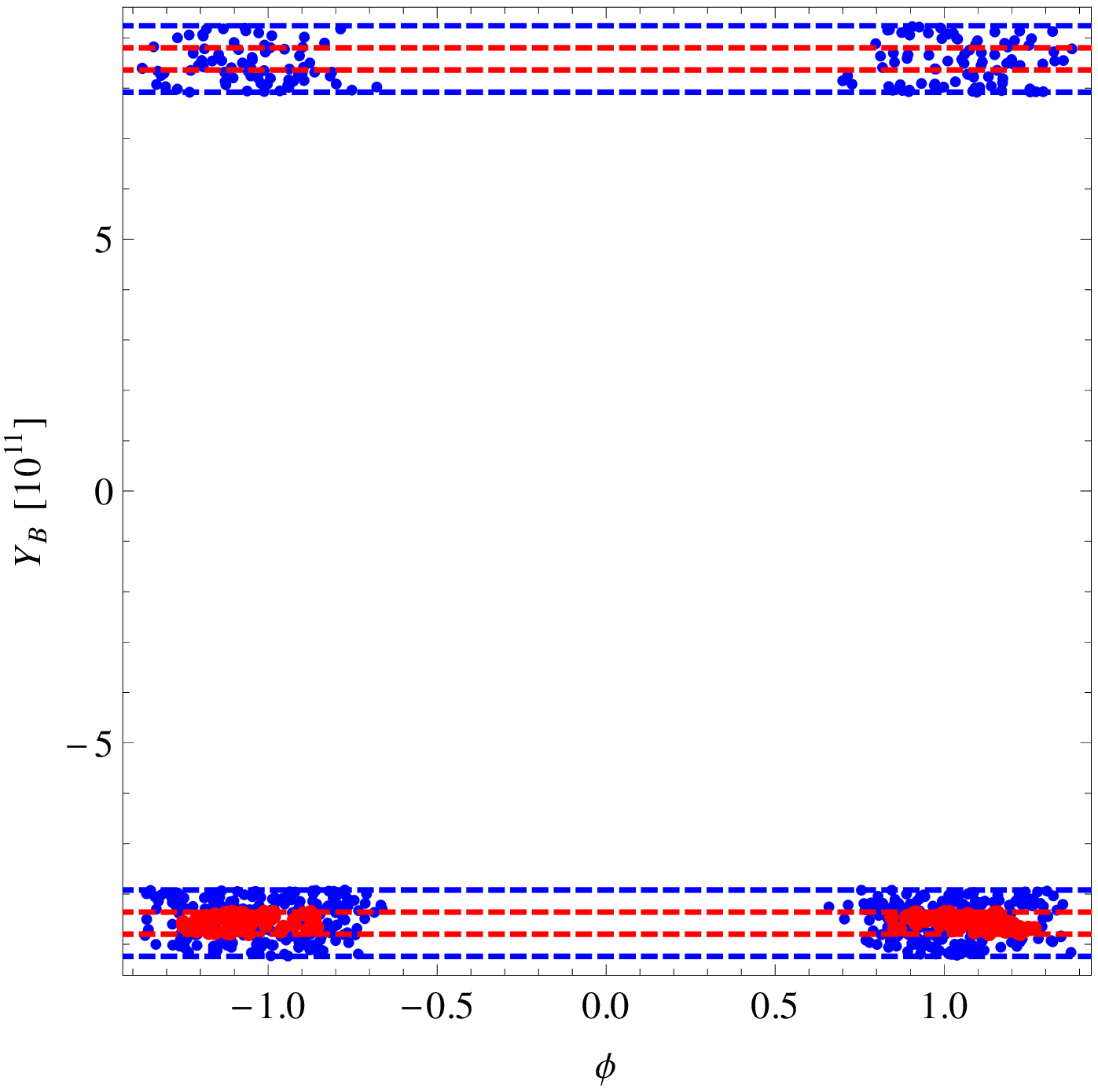} \\
\includegraphics[scale=0.49]{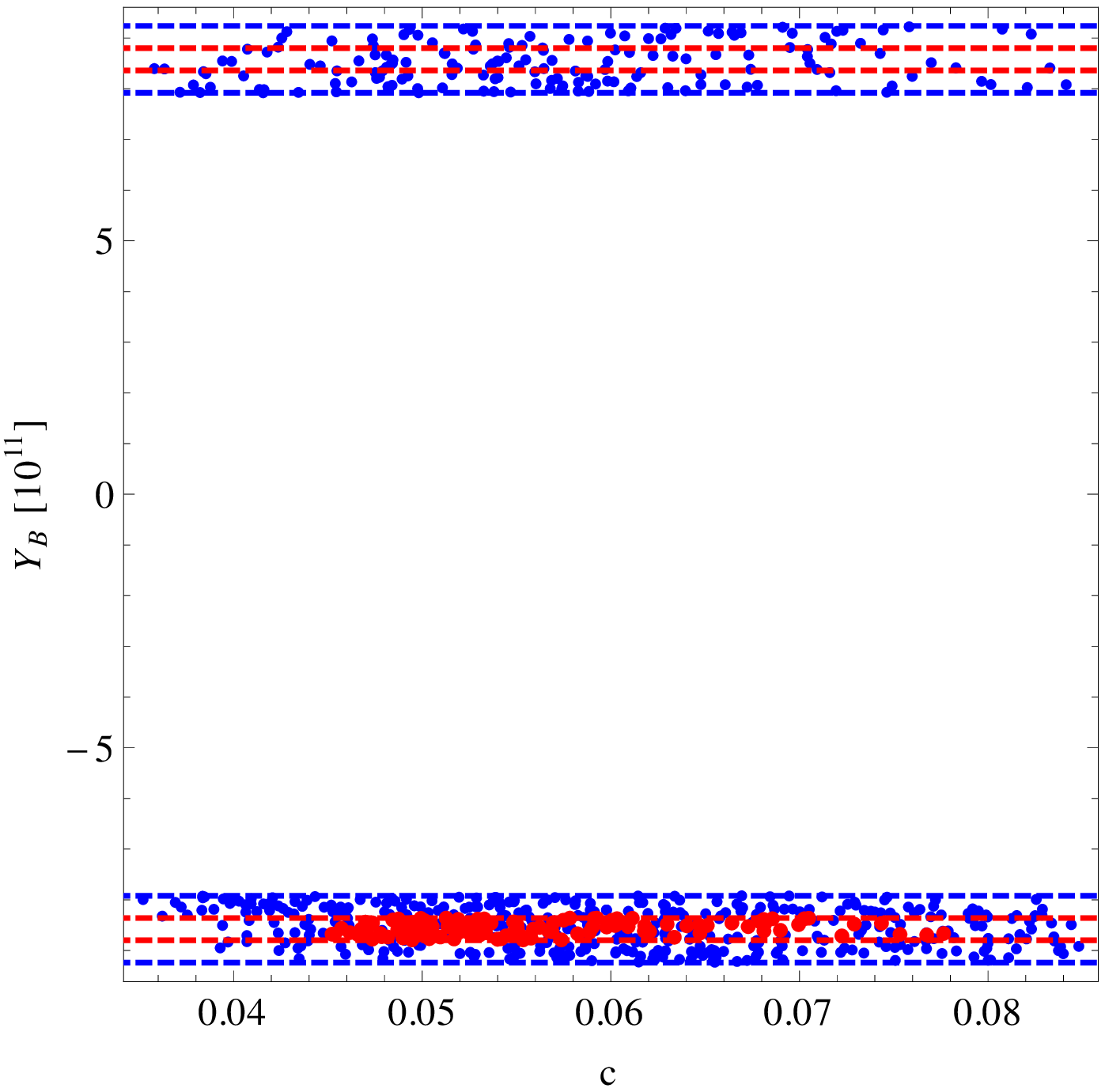} \hspace{0.7cm}
\includegraphics[scale=0.49]{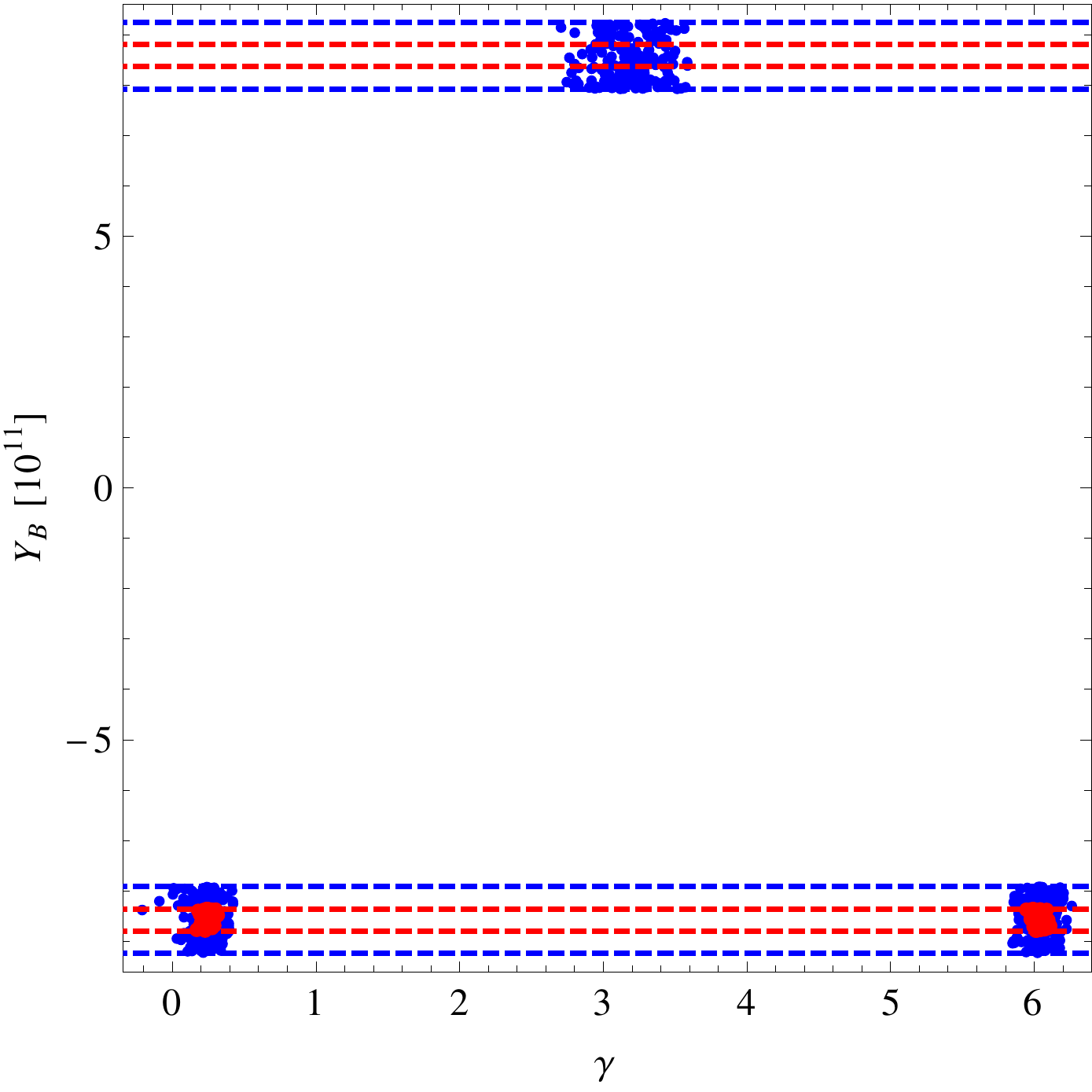} \\
\includegraphics[scale=0.49]{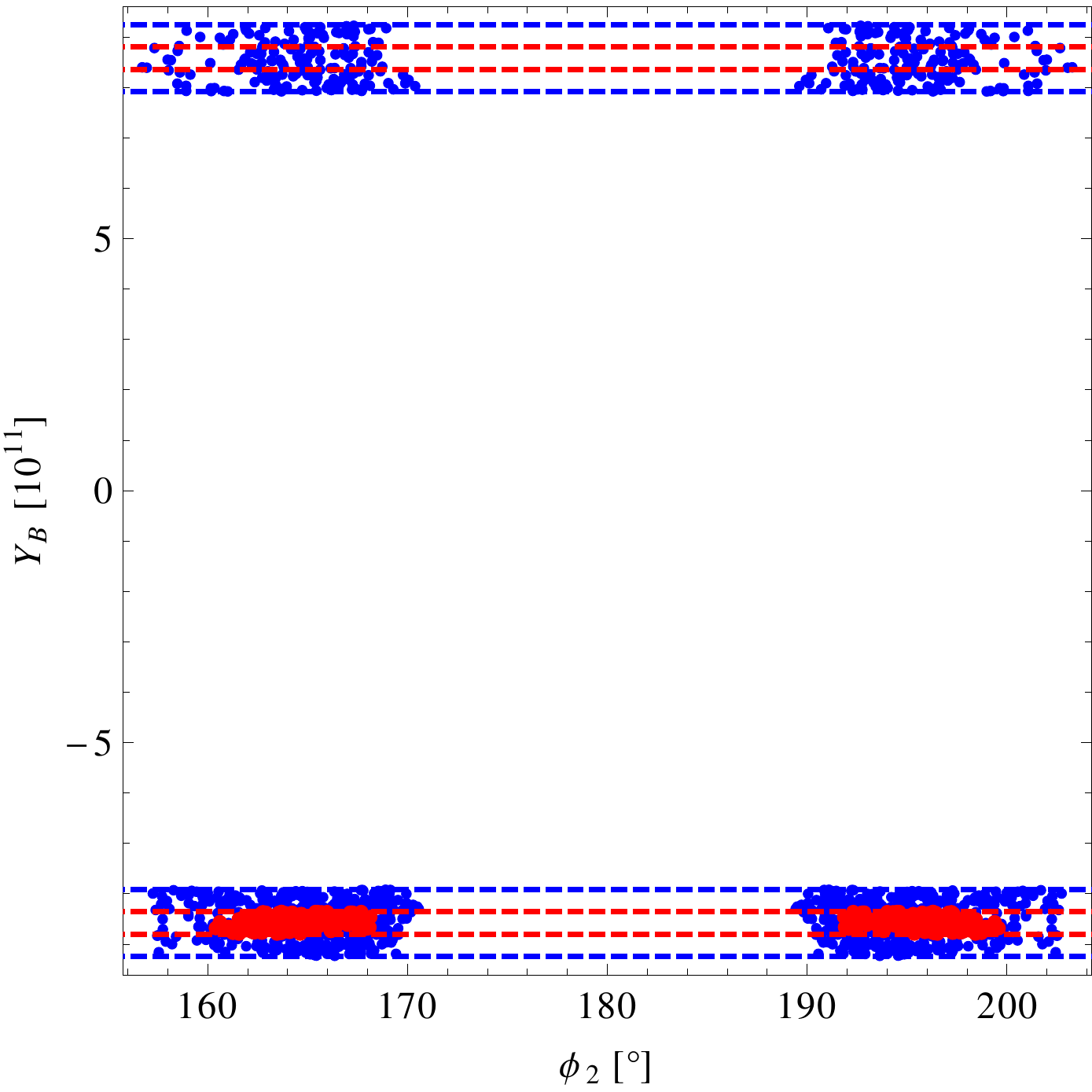} \hspace{0.7cm}
\includegraphics[scale=0.49]{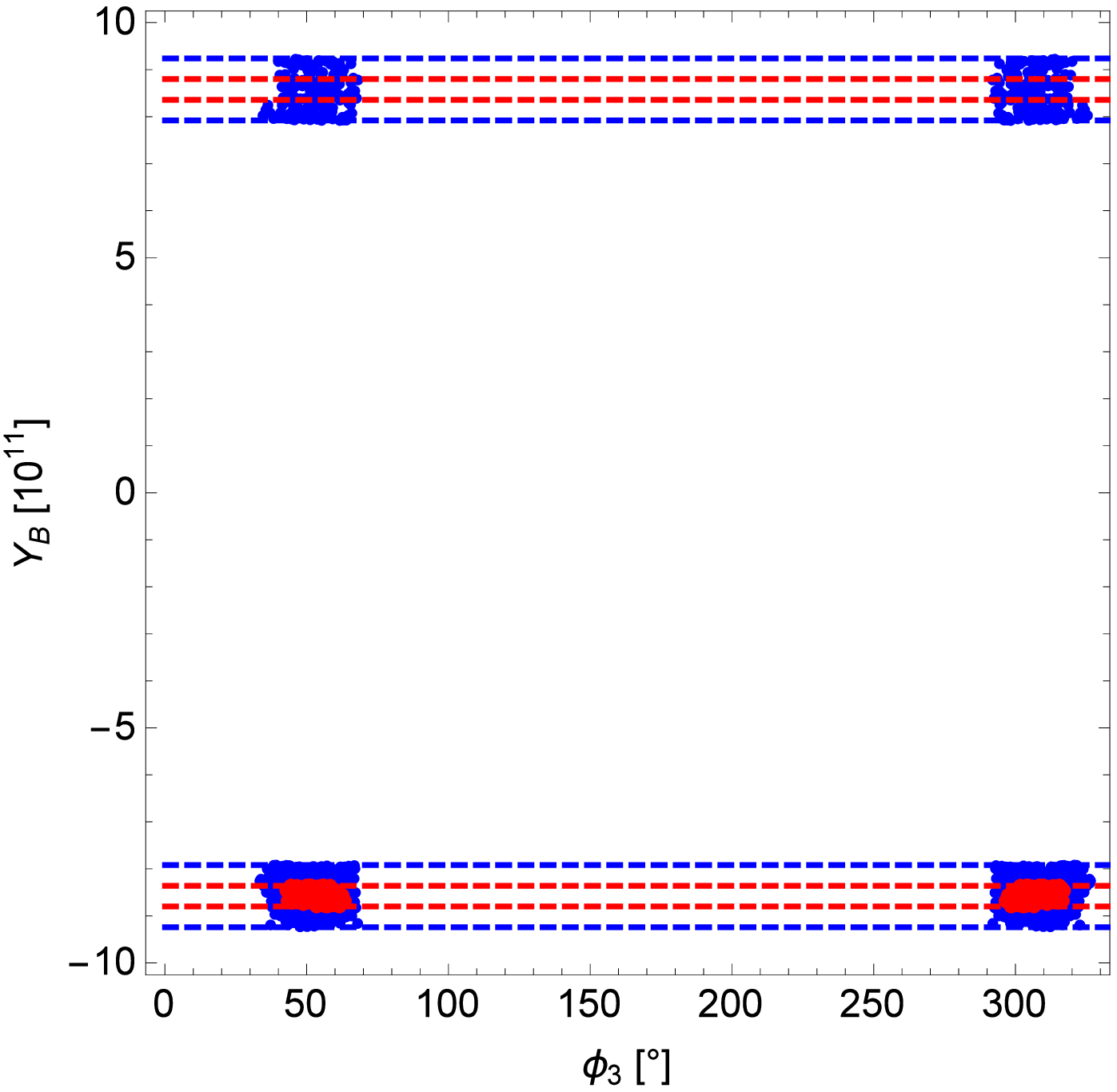} \\
\caption{
Results of our numerical scan for the total baryon asymmetry $Y_B$
in dependence of the six most relevant parameters.
Blue (red) points are in agreement within 3$\sigma$ (1$\sigma$) 
of the low energy
neutrino masses and mixings and $Y_B$ in our model.
}
\label{fig:YB_num}
\end{figure}

\begin{figure}
\centering
\includegraphics[scale=0.49]{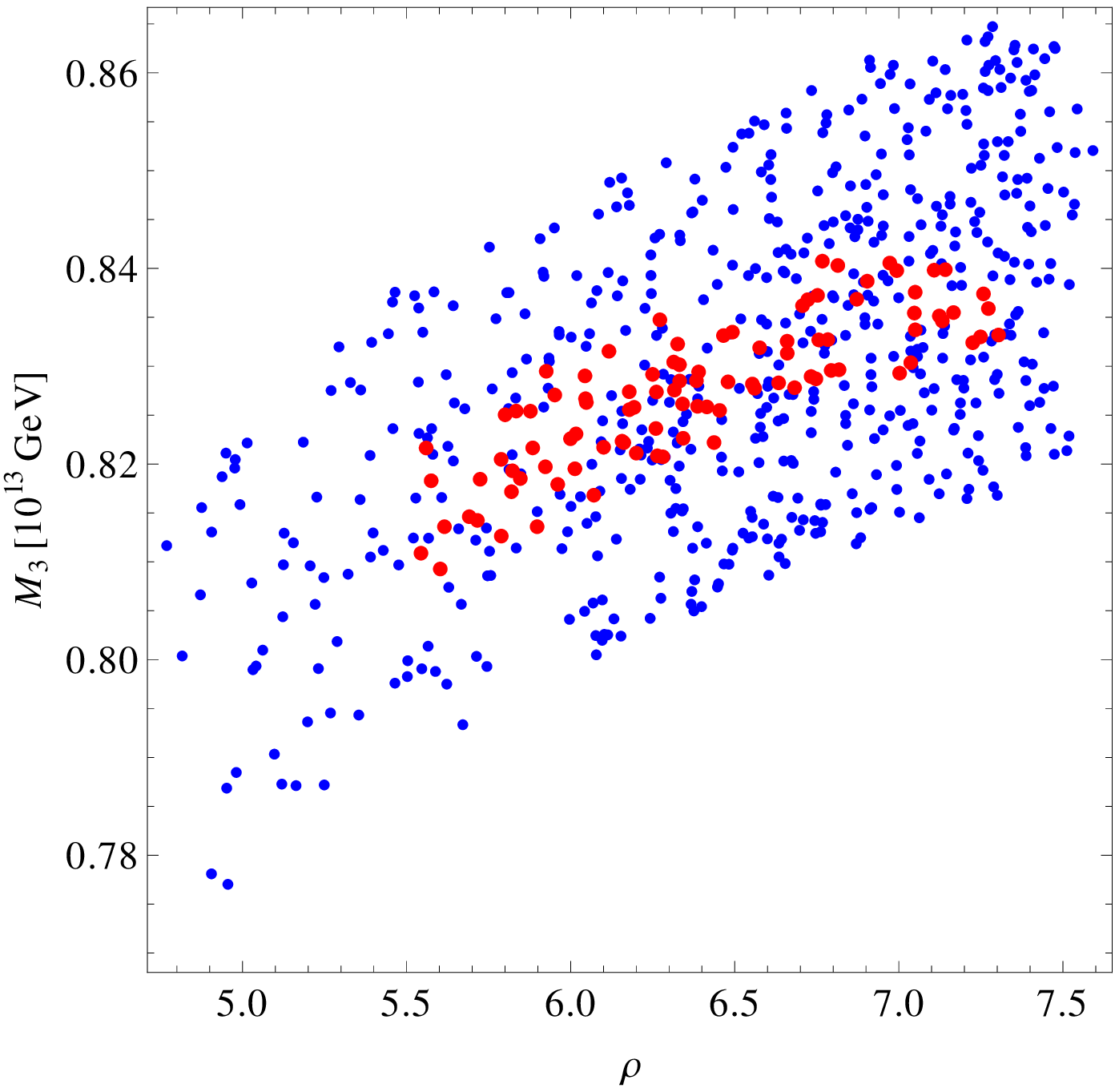} \hspace{0.7cm}
\includegraphics[scale=0.49]{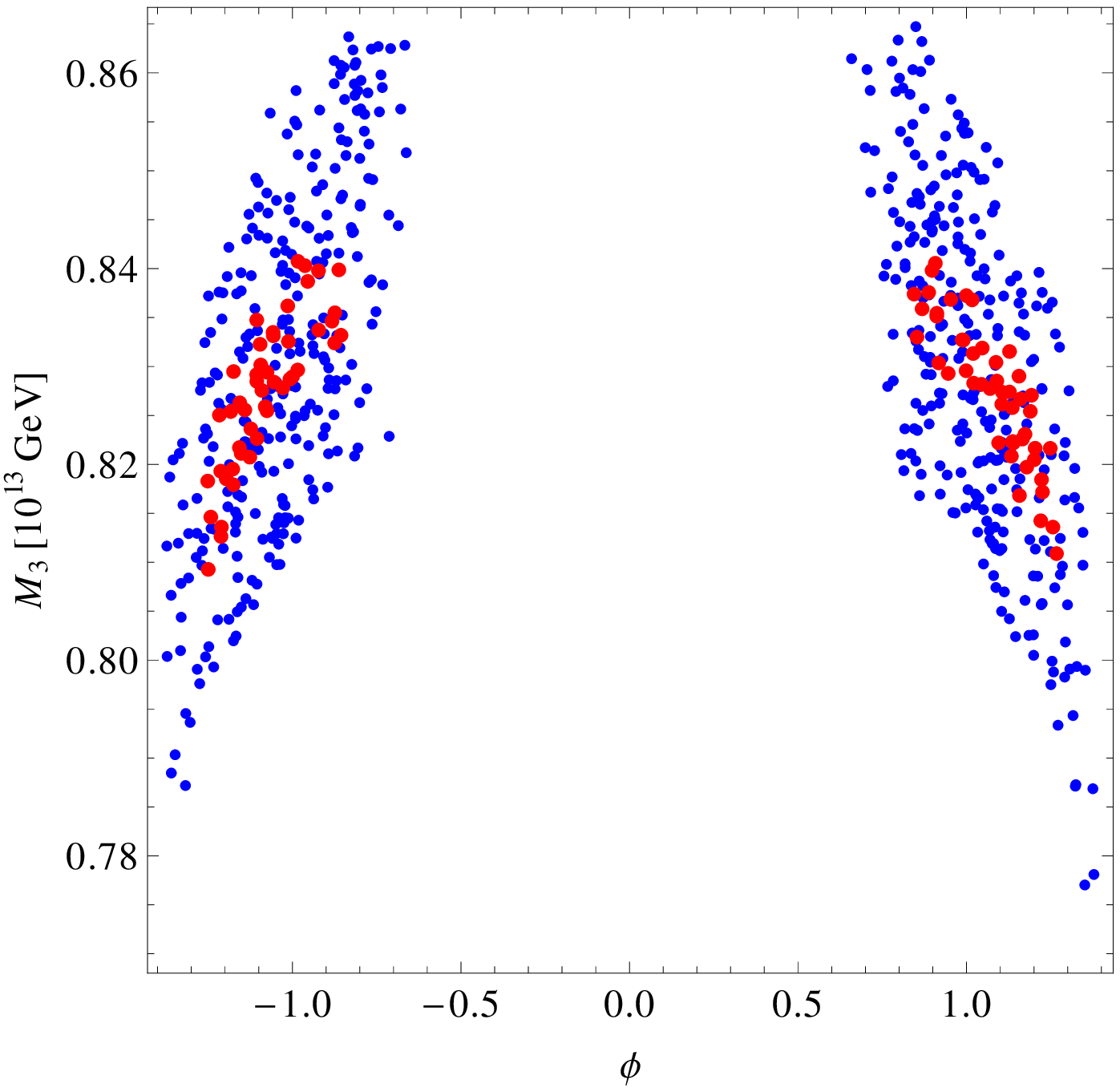} \\
\includegraphics[scale=0.49]{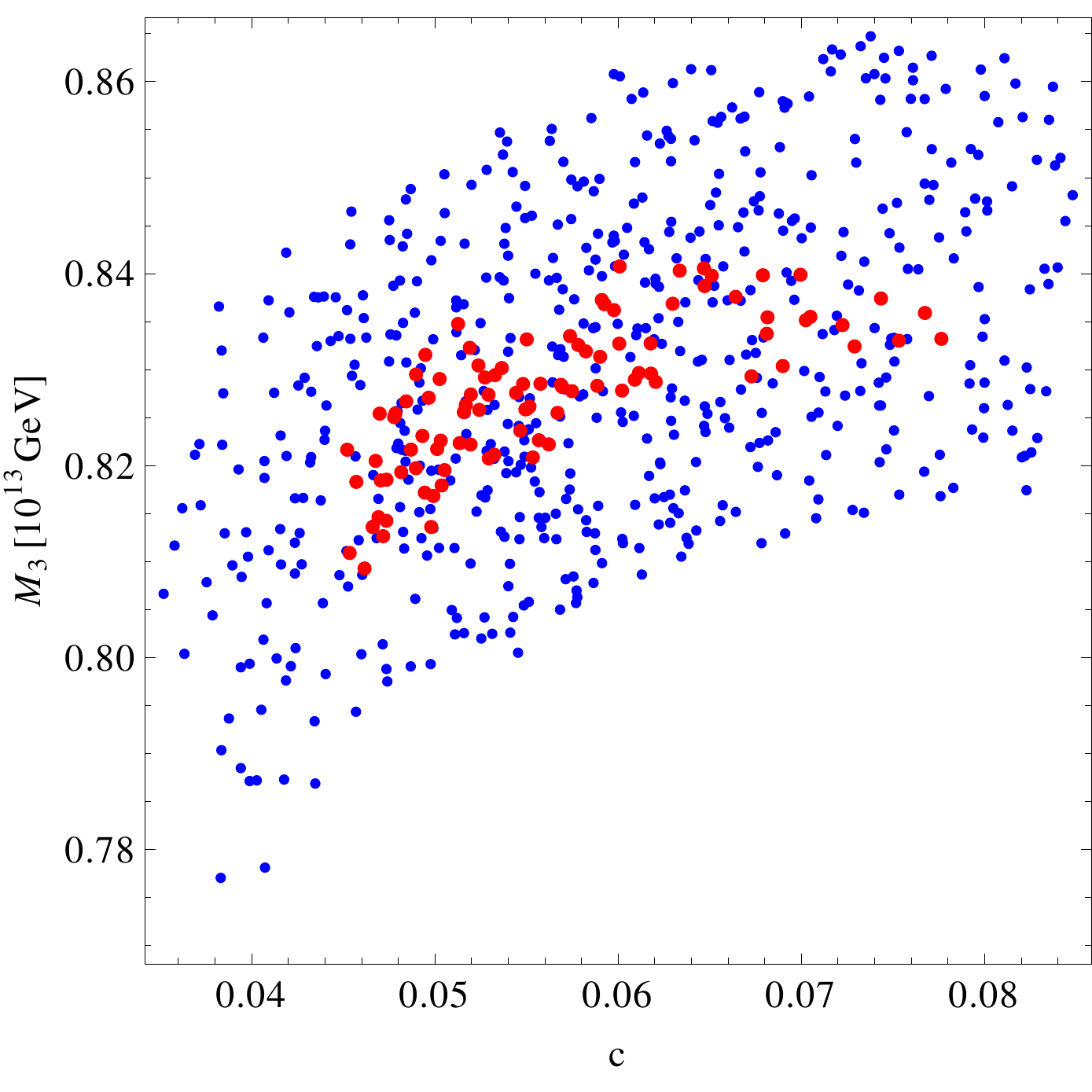} \hspace{0.7cm}
\includegraphics[scale=0.49]{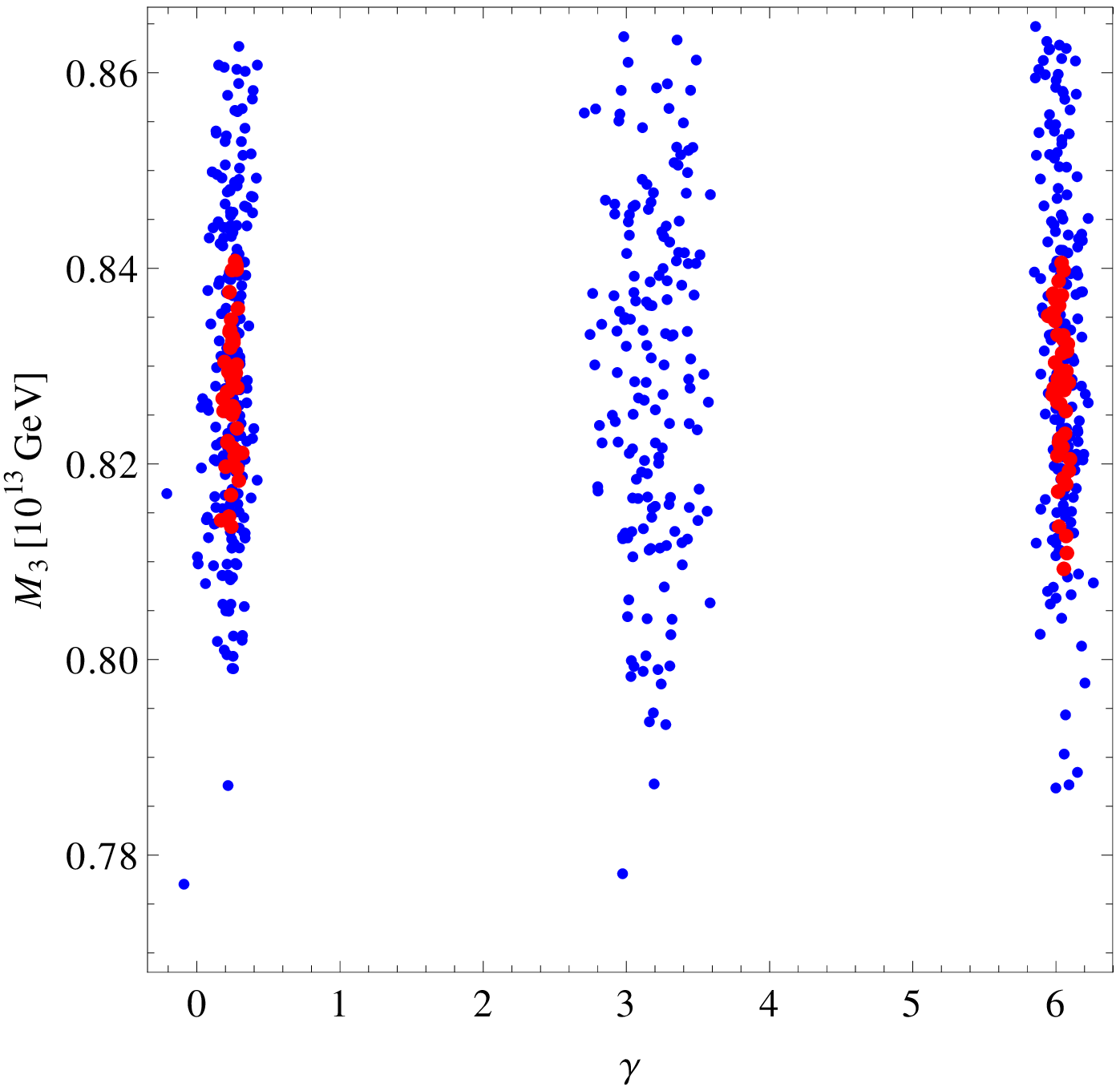} \\
\includegraphics[scale=0.49]{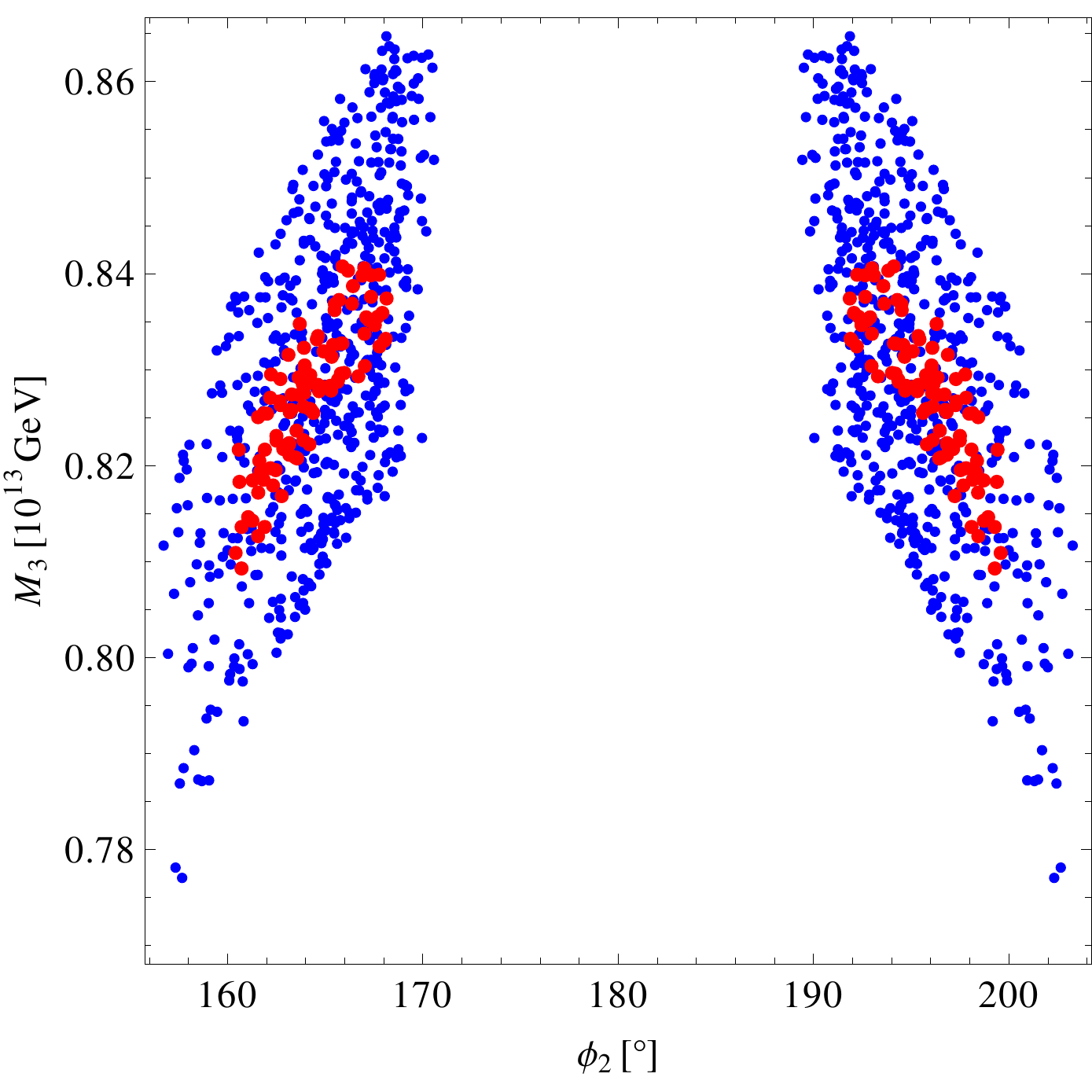} \hspace{0.7cm}
\includegraphics[scale=0.49]{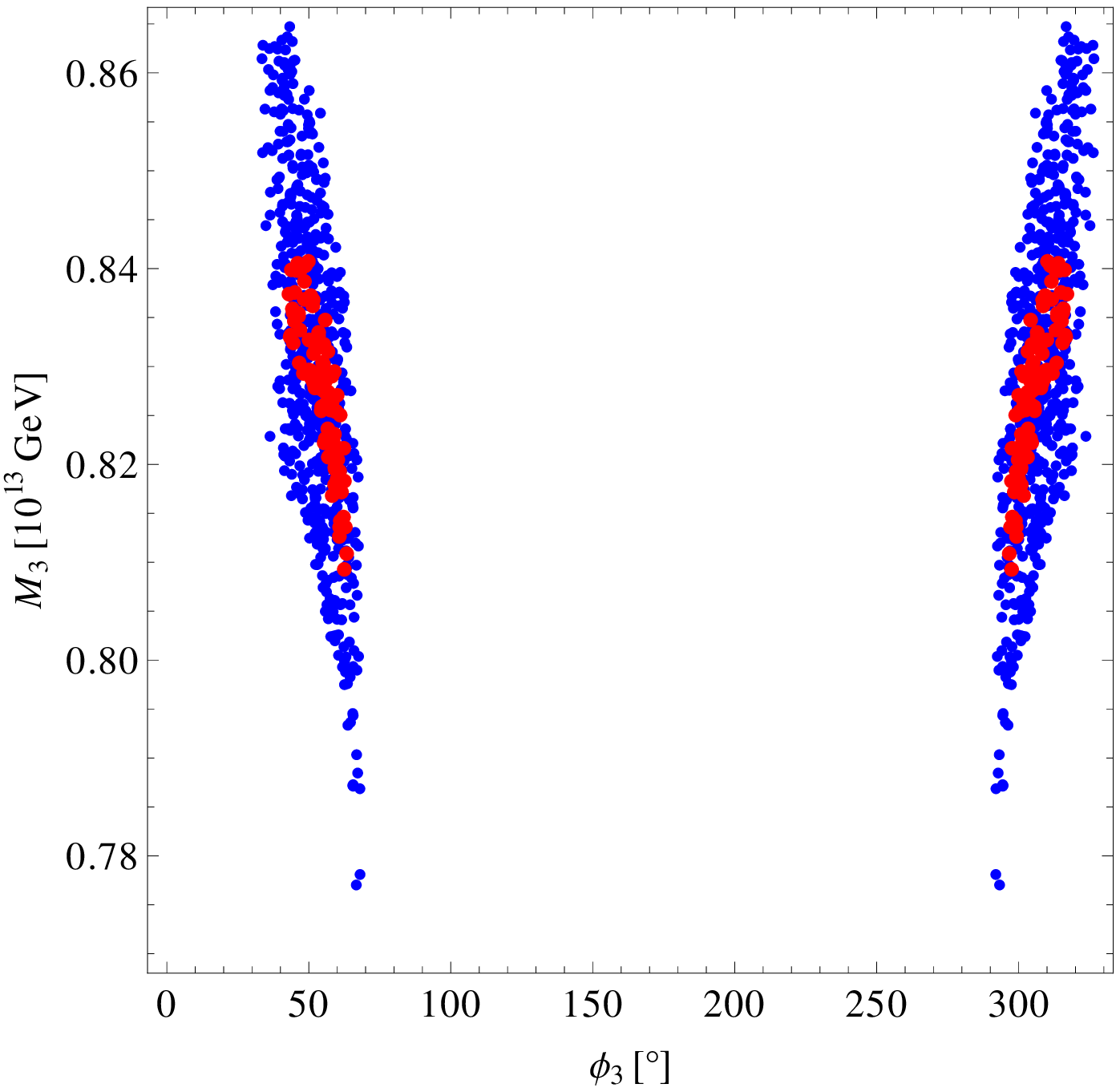} \\
\caption{
Results of our numerical scan showing the correlations
between the lightest right-handed neutrino mass and
the six most relevant parameters for leptogenesis.
Blue (red) points are in agreement within 3$\sigma$ (1$\sigma$) 
of the low energy
neutrino masses and mixings and $Y_B$ in our model.
}
\label{fig:M3_num}
\end{figure}

Moving on to our predictions for $Y_B$, we show the results of our parameter scan in Figs.~\ref{fig:YB_num} and \ref{fig:M3_num}.

We find parameter points which are in agreement with the 1$\sigma$ (3$\sigma$) range for $Y_B$. 
The ranges for the $\phi$ and $\rho$ are as in our previous study between $4.7$ and $7.6$ for $\rho$
and between $0.66$ and $1.38$ or -1.38 and -0.66 for $\phi$ since they are mostly determined by the
masses and mixing angles. However the constraints coming from the numerical solution of $Y_B$ now
forbid values of $\phi$ between -0.66 and 0.66.
Interesting is also the range for the mass of the lightest right-handed neutrino
which is in the range from $0.778 \cdot 10^{13}$~GeV to  $0.862 \cdot 10^{13}$~GeV which
shows a clear correlation to $c$ but no correlation to $\gamma$.The other correlations
can be seen in Fig.~\ref{fig:M3_num}. We furthermore want to note that the rather narrow
range of $M_3$ is due to the fact that we fixed $y_1^n$ to a certain value. As a rule of
thumb the scale of light neutrino masses has to remain the same and hence a variation of
10~\% in $y_1^n$ leads to a variation of 20~\% in $M_3$.

The size of the correction $c$ has to be in the range between 0.035 and 0.085 and $\gamma$ is in
narrow ranges around 0 or $\pi$. This also explains the disconnected regions
for the mixing angles and phases which did not appear so in our previous studies.
As in our previous study the sign of $Y_B$ is uniquely determined
by $\gamma$. For the 1$\sigma$ ranges of the parameters 
 the values
of $\gamma$ have to be in the region around 0. 
This implies that we only find parameter points for the
1$\sigma$ ranges compatible with a negative value of $Y_B$.
Note that this is mainly driven by the 1$\sigma$ range
of $\theta_{23}$ which is not very well determined. Indeed, with the 
results from the Valencian fitting collaboration \cite{Forero:2014bxa}
where $\theta_{23}$ is allowed to be in the second quadrant,
even at the 1$\sigma$ level positive values for $Y_B$ are possible.

As $\gamma$ is in narrow ranges around the CP conserving values, 
the Majorana phases are the major sources for CP violation here, 
as it can be seen from Fig.~\ref{fig:YB_num}.

\section{Summary and Conclusions}

In this work we have revised predictions for leptogenesis in an
SU(5)$\, \times \,$A$_{{5}}$ golden ratio GUT 
flavour model from \cite{Gehrlein:2015dxa}.
In that publication we had used approximations to calculate the baryon asymmetry which are
known to be precise only up to 20-30\%.
Instead we have solved here the full set of Boltzmann equations numerically and
could show that we can still successfully accommodate the experimental values of all
mixing angles, fermion masses and the baryon asymmetry even at the 1$\sigma$ level. 
Nevertheless, to do so we had to adjust some parameters. For instance, we have increased
the neutrino Yukawa coupling by 20\% from $y_1 = 0.1$ to $y_1 = 0.12$.

All of the main features of the original model are still valid. For instance, the neutrino mass sum rule
is correct up to small corrections and the inverted ordering is still ruled out. Furthermore,
the Yukawa coupling ratios $y_{\tau}/y_{b}\approx -3/2$ are unaffected by the modification of our
model and hence imply non-trivial constraints on the spectrum of the supersymmetric partners of
the Standard Model particles.

The phenomenological predictions are not changed significantly as well. For the lightest
neutrino mass we find the range $m_1 \in [0.012,0.022]$ eV, but the strong correlation between
$\theta_{13}$ and the phases is somewhat weakened due to an increased value of $c$.
For the phases we find $\delta \in[9 ^{\circ}, 119^{\circ}] ~\text{or}~ [239 ^{\circ}, 344^{\circ}]$,
$\alpha_1 \in [0^{\circ}, 134^{\circ}] ~\text{or}~ [220^{\circ}, 360^{\circ}]$ and $
\alpha_2 \in [66 ^{\circ}, 134^{\circ}] ~\text{or}~ [222^{\circ}, 282^{\circ}]$.
Related to the Majorana phases is the observable in neutrinoless double beta decay where
we predict $m_{ee}$ between 3.3 meV and 14.3 meV.

We have proposed the first SU(5)$\, \times \,$A$_{5}$ model to our knowledge which can
successfully accommodate for all parameters in the neutrino sector and for the baryon asymmetry
of the Universe as calculated from the Boltzmann equations and for the not yet measured quantities
in the neutrino sector we make testable predictions.
For this our model needs comparatively few parameters which makes it very appealing and testable in
future experiments.

\section*{Acknowledgements}
X.~Z. would like to thank E. Molinaro for sharing 
his code on solving Boltzmann equations. S.T.~P. acknowledges very 
useful discussions with E. Molinaro.                        
This work was supported in part  by the European Union FP7
ITN INVISIBLES (Marie Curie Actions, PITN-GA-2011-289442-INVISIBLES),
by the INFN program on Theoretical Astroparticle Physics (TASP),
by the research grant  2012CPPYP7 ({\sl  Theoretical Astroparticle Physics})
under the program  PRIN 2012 funded by the 
Italian MIUR and by the World Premier International Research Center
Initiative (WPI Initiative), MEXT, Japan (STP). 
J.~G. acknowledges support by the DFG-funded research training group GRK 1694 ``Elementarteilchenphysik bei h\"ochster Energie und h\"ochster Pr\"azision''.

\end{document}